\documentclass[10pt, conference, letterpaper]{IEEEtran}
\usepackage[noadjust]{cite}

\usepackage{graphicx}
\usepackage{amsmath}
\usepackage{pgfplots}
\usepackage{algorithm}
\usepackage{algorithmic}
\usepackage{subfigure}
\usepackage{amssymb}
\usepackage{amsthm}
\usepackage{tikz}
\usetikzlibrary{patterns, shapes.geometric, arrows, arrows.meta, calc}
\usepackage{color,soul}
\usepackage{multirow}
\usepackage[nodisplayskipstretch]{setspace}
\begin{document}
	
	\title{A Hierarchical Framework of Cloud Resource Allocation and Power Management Using Deep Reinforcement Learning}
	\author{\IEEEauthorblockN{Ning Liu$^*$, Zhe Li$^*$, Jielong Xu$^*$, Zhiyuan Xu, Sheng Lin, Qinru Qiu, Jian Tang, Yanzhi Wang}
			\IEEEauthorblockA{Department of Electrical Engineering and Computer Science,\\ Syracuse University, Syracuse, NY 13244, USA\\
			\{nliu03, zli89, zxu105, jxu21, shlin, qiqiu, jtang02, ywang393\}@syr.edu}
			
			}
	\maketitle
	\begin{abstract}
Automatic decision-making approaches, such as reinforcement learning (RL), have been applied to (partially) solve the resource allocation problem adaptively in the cloud computing system.
However, a complete cloud resource allocation framework exhibits high dimensions in state and action spaces, which prohibit the usefulness of traditional RL techniques. 
In addition, high power consumption has become one of the critical concerns in design and control of cloud computing systems, which degrades system reliability and increases cooling cost. 
An effective dynamic power management (DPM) policy should minimize power consumption while maintaining performance degradation within an acceptable level.
Thus, a joint virtual machine (VM) resource allocation and power management framework is critical to the overall cloud computing system. Moreover, novel solution framework is necessary to address the even higher dimensions in state and action spaces.

In this paper, we propose a novel hierarchical framework for solving the overall resource allocation and power management problem in cloud computing systems. The proposed hierarchical framework comprises a global tier for VM resource allocation to the servers and a local tier for distributed power management of local servers. 
The emerging deep reinforcement learning (DRL) technique, which can deal with complicated control problems with large state space, is adopted to solve the global tier problem. 
Furthermore, an autoencoder and a novel weight sharing structure are adopted to handle the high-dimensional state space and accelerate the convergence speed. On the other hand, the local tier of distributed server power managements comprises an LSTM based workload predictor and a model-free RL based power manager, operating in a distributed manner. 
Experiment results using actual Google cluster traces show that our proposed hierarchical framework significantly saves the power consumption and energy usage than the baseline while achieving no severe latency degradation. Meanwhile, the proposed framework can achieve the best trade-off between latency and power/energy consumption in a server cluster.
	\end{abstract}
    

	\section{Introduction}
	\label{Sec:intro}
    \let\thefootnote\relax\footnote{\vskip -3em \noindent\rule{\columnwidth}{0.4pt}\\\noindent $^*$Ning Liu, Zhe Li and Jielong Xu contributed equally to this work}
    Cloud computing has emerged as the most popular computing paradigm in today's computer industry. Cloud computing with virtualization technology enables computational resources (including CPU, memory, disk, communication bandwidth, etc.) in data centers or server clusters to be shared by allocating virtual machines (VMs) in an on-demand manner. The effective adoption of VMs in data centers is one of the key enablers of the large-scale cloud computing paradigm. To support such a feature, an efficient and robust scheme of VM allocation and management is critical. Due to the time-variance of workloads~\cite{mishra2010towards,khan2012workload}, it is desirable to perform cloud resource allocation and management in an online adaptive manner.    
	Automatic decision-making approaches, such as \emph{reinforcement learning} (RL)~\cite{sutton1998reinforcement}, have been applied to (partially) solve the resource allocation problem in the cloud computing system~\cite{Dutreilh11,Barrett2013,Galstyan04}.
	Key properties of RL-based methods are suitable for the cloud computing systems because they do not require \emph{a priori} modeling of state transition, workload, and power/performance of the underlying system, i.e., they are essentially \emph{model-free}. 
	Instead, the RL-based agents learn the optimal resource allocation decision and control the system operation in an online fashion as the system runs.
    
    However, a complete resource allocation framework in the cloud computing systems exhibits high dimensions in state and action spaces. 
    For example, a \emph{state} in the state space may be the Cartesian product of characteristics and current resource utilization level of each server (for hundreds of servers) as well as current workload level (number and characteristics of VMs for allocation). 
    An \emph{action} in the action space may be the allocation of VMs to the servers (a.k.a. physical machines) and allocating resources in the servers for VM execution. 
    The high dimensions in state and action spaces prohibit the usefulness of traditional RL techniques to the overall cloud computing system in that the convergence speed of traditional RL techniques is in general proportional to the number of state-action pairs~\cite{sutton1998reinforcement} and will be impractically long with high state and action dimensions. 
    As a result, previous works only used RL to dynamically adjust the power status of a single physical server~\cite{tesauro2007managing} or the number of homogeneous VMs to an application~\cite{lin2016reinforcement,Barrett2013}, in order to restrict the state and action spaces.
    
    In addition, power consumption has become one of the critical concerns in design and control of cloud computing systems. High power consumption degrades system reliability and increases the cooling cost for high-performance systems. Dynamic power management (DPM), defined as the selective shut-off or slow-down of system components that are idle or underutilized, has proven to be an effective technique for reducing power dissipation at system level~\cite{benini2000survey}. 
    An effective DPM policy should minimize power consumption while maintaining performance degradation to an acceptable level. The design of such DPM policies has been an active research area, while a variety of adaptive power management techniques including machine learning have been applied \cite{benini2000survey, dhiman2006dynamic,jung2007dynamic}. The DPM policies have mainly been applied to the server level, and therefore depend on the results of VM resource allocations. Therefore, a joint VM resource allocation and power management framework, which targets at the whole data center or server cluster, would be critical to the overall cloud computing system. And obviously, the joint management framework exhibits even higher dimensions in state and action spaces and requires a novel solution framework.
    
    
    Recent breakthroughs of \emph{deep reinforcement learning} (DRL) in AlphaGo and playing Atari set a good example in handling large state space of complicated control problems~\cite{silver2016mastering,mnih2013playing}, and could be potentially utilized to solve the overall cloud resource allocation and power management problem. The convolutional neural network, one example of \emph{deep neural networks} (DNN), was used to effectively extract useful information from high-dimensional image input and build a correlation between each state-action pair $(s,a)$ and the associated \emph{value function} $Q(s,a)$, which is the expected accumulative rewards (or costs) that the agent aims to maximize (or minimize) in RL. 
    For online operations, a deep Q-learning framework was also proposed to derive the optimal action $a$ at each state $s$ in order to maximize (or minimize) the corresponding $Q(s,a)$ value. 
    Although promising, both the DNN and the deep Q-learning framework need to be modified for applications in cloud resource allocation and power management.
    Moreover, the DRL framework requires a relatively low-dimensional action space~\cite{mnih2013playing} because in each decision epoch the DRL agent needs to enumerate all possible actions at current state and perform inference using DNN to derive the optimal $Q(s,a)$ value estimate, which implies that the action space in cloud resource allocation and power management needs to be significantly reduced.
    
    To fulfill the above objectives, in this paper we propose a novel \emph{hierarchical framework} for solving the overall resource allocation and power management problem in cloud computing systems. The proposed hierarchical framework comprises a \emph{global tier} for VM resource allocation to the servers and a \emph{local tier} for power management of local servers. Besides the enhanced scalability and reduced state/action space dimensions, the proposed hierarchical framework enables to perform the local power managements of servers \emph{in an online and distributed manner}, which further enhances the parallelism degree and reduces the online computational complexity.
    
    The global tier of VM resource allocation exhibits large state and action spaces, and thus the emerging DRL technique is adopted to solve the global tier problem. In order to significantly reduce the action space, we adopt a continuous-time and event-driven decision framework in which each decision epoch coincides with the arrival time of a new VM request. In this way the action at each decision epoch is simply the target server for VM allocation, which ensures that the available actions are enumerable. Furthermore, an \textit{autoencoder} \cite{bengio2009learning} and a novel \emph{weight sharing structure} are adopted to handle the high-dimensional state space and accelerate the convergence speed, making use of the specific characteristics of cloud computing systems. On the other hand, the local tier of server power managements comprises a \textit{workload predictor} and a \textit{power manager}. The workload predictor is responsible for providing future workload predictions to facilitate the power management algorithm, and we adopt the \emph{long short-term memory} (LSTM) network \cite{hochreiter1997long} due to its ability to capture long-term dependencies in time-series prediction. Based on the workload prediction results and the current information, the power manager adopts the model-free RL technique to adaptively determine the suitable action for turning ON/OFF of the servers in order for simultaneous reductions of power/energy consumption and job (VM) latency.
    
    Experiment results using actual Google cluster traces \cite{clusterdata:Reiss2011} show that proposed hierarchical framework significantly save the power consumption/energy usage than the baseline while achieves similar average latency. In a 30-server cluster, with $95,000$ job requests, the proposed hierarchical framework can save 53.97\% power and energy consumptions. Meanwhile, the proposed framework can achieve the best trade-off between latency and power/energy consumption in a server cluster. In the same case, the proposed framework gives the average per-job latency saving with the same energy usage up to 16.16\%, and the average power/energy saving with the same latency up to 16.20\%.
       
	\section{Background of the Agent-Environment Interaction System and Continuous-Time Q-Learning}
    \label{Sec:background}

\subsection{Agent-Environment Interaction System}
\begin{figure}
\centering
\includegraphics[width=0.85\columnwidth]{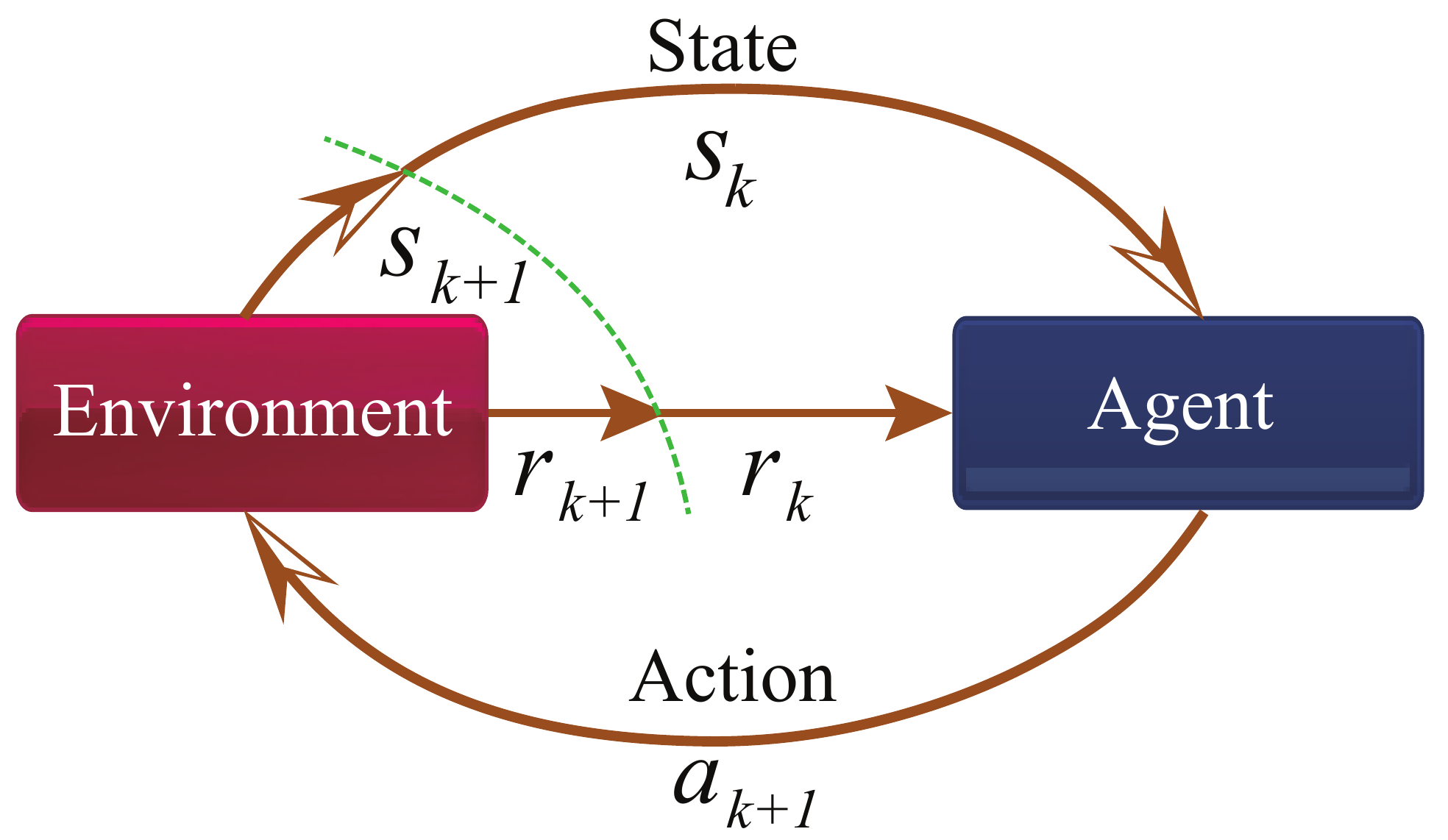}
\caption{Illustration of the agent-environment interaction system.}
\label{Fig:RLmodel}
\end{figure}
As shown in Fig.~\ref{Fig:RLmodel}, the general agent-environment interaction modeling (of both traditional RL and the emerging DRL) consists of an agent, an environment, a finite state space $S$, a set of available actions $A$, and a reward function: $S\times A\rightarrow R$. 
The decision maker is called the \emph{agent}, and should be trained as the interaction system runs.
The agent needs to interact with the outside, which is called the \emph{environment}. The interaction between the agent and the environment is a continual process. 
At each decision epoch $k$, the agent will make decision $a_k$ based on the current state $s_k$ of the environment. 
Once the decision is being made, the environment would receive the decision and make corresponding changes, and the updated new state $s_{k+1}$ of the environment would be presented to the agent for making future decisions. 
The environment also provides reward $r_k$ to the agent depending on the decision $a_k$, and the agent tries to maximize some notion of the cumulative rewards over time. 
This simple reward feedback mechanism is required for the agent to learn its optimal behavior and policy.

\subsection{Continuous-Time Q-Learning for Semi-Markov Decision Process (SMDP)}
In the Q-learning procedure \cite{watkins1992q}, a representative algorithm in general RL, the agent aims to maximize a \emph{value function} $Q(s,a)$, which is the expected accumulated (with discounts) reward function when system starts at state $s$ and follows action $a$ (and certain policy thereafter). $Q(s,a)$ is given as:
\begin{equation}
Q(s,a)=\textbf{E}\bigg[\int_{t_0}^{\infty}e^{-\beta(t-t_0)}r(t)dt\bigg|s_0=s, a_0=a\bigg]
\end{equation}
for continuous-time systems where $r(t)$ is the reward rate function and $\beta$ is the discount rate. The $Q(s,a)$ for discrete-time systems can be defined similarly.

The Q-learning for SMDP is an online adaptive RL technique that operates in continuous time domain in an event-driven manner \cite{duff1995reinforcement}, which could reduce the overheads associated with periodic value updates in discrete-time RL techniques. Please note that the name of the technique has the term ``SMDP" ONLY because it is proven to achieve the optimal policy under SMDP environment. In fact, it could be utilized to non-stationary environments as well with excellent results~\cite{wang2011deriving}. In Q-learning for SMDP, the definition of value function $Q(s,a)$ is continuous-time based and given in Eqn. (1). At each decision epoch $t_k$, the RL agent selects the action $a_k$ using certain policy, e.g., the $\epsilon$-greedy policy \cite{sutton1996generalization}, similar to discrete-time RL techniques. At the next decision epoch $t_{k+1}$ triggered by state transition, the value updating rule (from the $k$-th estimate to the $(k+1)$-th estimate) is given by the following:
\begin{align}
 &Q^{(k+1)}(s_k,a_k)\leftarrow Q^{(k)}(s_k,a_k)+\alpha \cdot
\Big(\frac{1-e^{-\beta \tau_k}}{\beta}r(s_k,a_k)+\nonumber\\
&\max_{a'}e^{-\beta \tau_k}Q^{(k)}(s_{k+1},a')-Q^{(k)}(s_k,a_k)\Big)
\end{align}
where $Q^{(k)}(s_k,a_k)$ is the value estimate at decision epoch $t_k$, $r(s_k,a_k)$ is the reward function, $\tau_k$ is the sojourn time that system remains in state $s_k$ before a transition occurs, $\alpha\leq 1$ is the \emph{learning rate}, and $\beta$ is \textit{discount rate}.
    \section{System Model and Problem Statement}
    \label{Sec:problem}
\begin{figure}
\centering
\includegraphics[width=0.9\columnwidth]{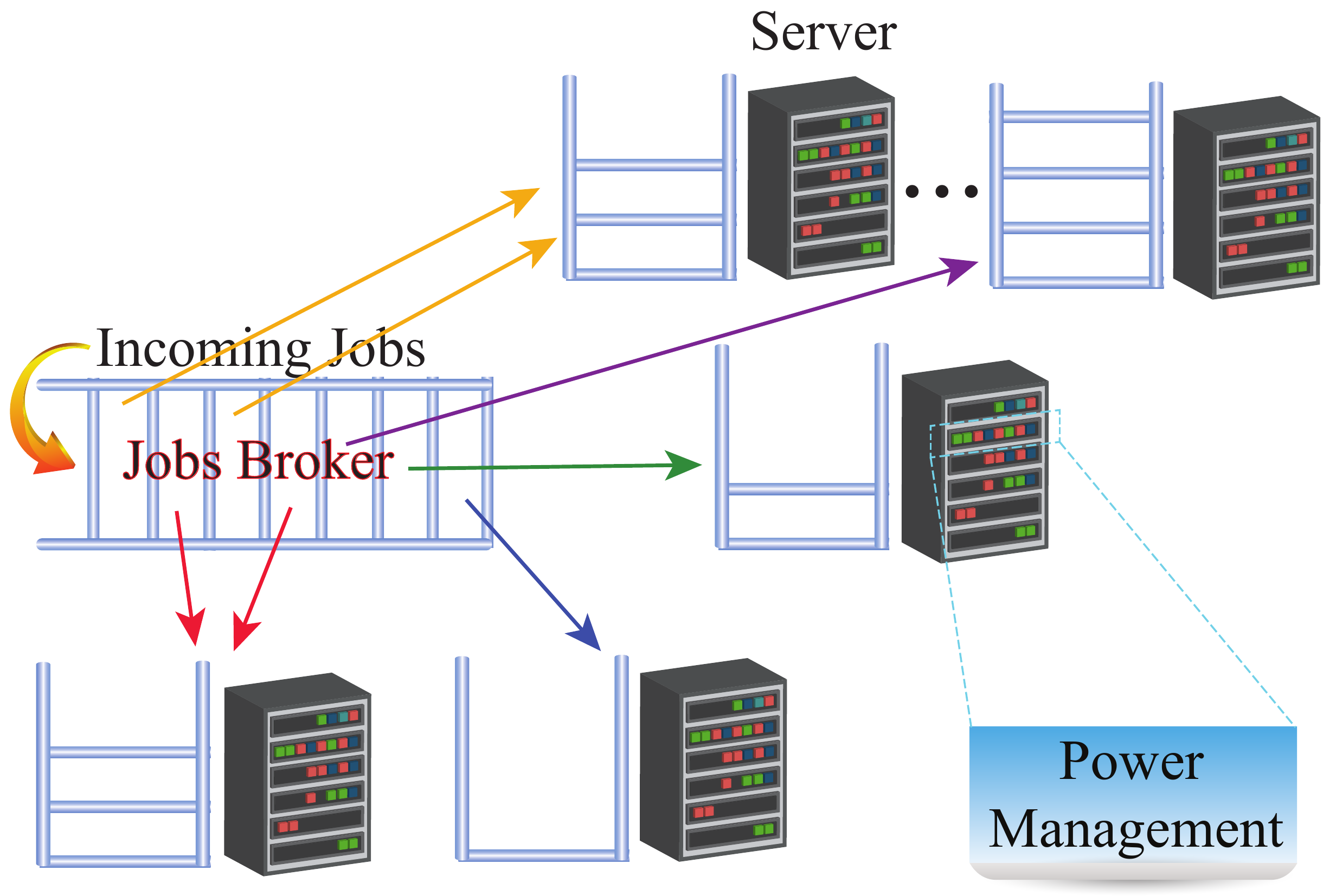}
\caption{Cloud resource allocation and power management framework, comprising both the global tier and the local tier.}
\label{Fig:cluster}
\end{figure}

We consider a server cluster\footnote{This work can be generally applied to multiple server clusters or an overall data center.} with $M$ physical servers that offer $D$ types of resources with respect to the cloud resource allocation and power management framework in this paper.
Usually, a server can be in active mode or sleep mode for power saving.
We denote $\mathcal{M}$ as the set of physical servers, and $\mathcal{D}$ as the set of resources. Fig. 2 provides an illustration of the hierarchical cloud resource allocation and power management framework, comprising a global tier and a local tier. 
A job broker, controlled by the global tier in the proposed hierarchical framework, dispatches jobs that request resource for processing to one of the servers in the cluster at its arrival time, as shown in Fig. \ref{Fig:cluster}.
Each server queues all assigned jobs and allocates resources for them in a \textit{first-come-first-serve (FCFS)} manner.
If a server has insufficient resource to process a job, it waits until sufficient resource is released by completed jobs. On the other hand, the local tier performs power management and turns ON/OFF of each server, in a distributed manner. Both the job scheduling and the local power management will significantly affect the overall power consumption and performance of the server cluster.

\begin{figure}
\centering
\begin{tikzpicture}[scale=0.9]
	\begin{axis}[xmin=0,xmax=15,ymin=0,ymax=11,
			axis line style={very thick},
	        axis y line = left, ylabel=\bf{CPU(\%)},
	        ylabel near ticks,
	        xlabel near ticks,
	        axis x line = bottom, xlabel=\bf{time},
	        x label style = {at={(axis cs:16, 1)}},
	        unit vector ratio=1 0.75 1,
	        xtick=\empty,
	        ytick={2.5, 5, 7.5, 10},
	        yticklabels = {$25$, $50$, $75$, $100$},
	        clip=false]
		\draw[thick] (axis cs:1,0) rectangle (axis cs:7.5,5) node[midway]
			{\begin{tabular}{c} Job 1\\latency: $t_4 - t_1$\end{tabular}};
		\draw[thick] (axis cs:3,5) rectangle (axis cs:11,9) node[midway]
			{\begin{tabular}{c} Job 2\\latency: $t_5 - t_2$\end{tabular}};
		\draw[thick] (axis cs:7.5,0) rectangle (axis cs:14,4) node[midway]
			{\begin{tabular}{c} Job 3\\latency: $t_6 - t_3$\end{tabular}};
		\draw[dashed, latex-, thick] (axis cs:1,5) -- (axis cs:1,10.5) node[above] {$t_1$};
		\draw[dashed, latex-, thick] (axis cs:3,9) -- (axis cs:3,10.5) node[above] {$t_2$};
		\draw[dashed, latex-, thick] (axis cs:6,9) -- (axis cs:6,10.5) node[above] {$t_3$};
		\draw[dashed, latex-, thick] (axis cs:7.5,9) -- (axis cs:7.5,10.5) node[above] {$t_4$};
		\draw[dashed, latex-, thick] (axis cs:11,9) -- (axis cs:11,10.5) node[above] {$t_5$};
		\draw[dashed, latex-, thick] (axis cs:14,4) -- (axis cs:14,10.5) node[above] {$t_6$};
	\end{axis}
\end{tikzpicture}
\caption{An example of job execution on a server when the server is active. 
	}
\label{Fig:dispatch}
\end{figure}
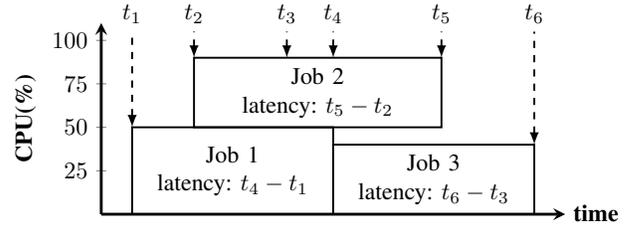

We demonstrate how jobs are executed on an active server with only CPU resource usage in Fig. \ref{Fig:dispatch} as an example.
Job $1$ consumes $50\%$ of CPU, while each of job $2$ and $3$ requires $40\%$.
Job $1$, $2$ and $3$ arrive at $t_1$, $t_2$ and $t_3$, respectively, and
complete at $t_4$, $t_5$ and $t_6$, respectively. 
We assume that the server is in active mode at time $0$. When job $1$ and $2$ arrive, there are enough CPU resources so their requirements are satisfied immediately. 
When job $3$ arrives, it waits until the job $1$ is completed, and the waiting time is $t_4 - t_3$.
The \textit{job latency} is defined as the duration between the job arrival and completion.
Therefore the latency of job $3$ is $t_6 - t_3$, which is longer than the
job duration. 
To reduce the job latency, the job broker should avoid overloading servers.
A scheduling scheme must be developed for dynamically assigning the jobs to servers and allocating resources in each server.

When a job is assigned to a server in sleep mode, it takes $T_\textbf{on}$ time to change the server into the active mode. 
Similarly, it takes $T_\textbf{off}$ time to change the server back to the sleep mode, which decision is made by the local tier of power manager (in a distributed manner).
We assume the power consumption of a server in the sleep mode is zero, and the power consumption of a server at time $t$ in the active mode
is a function of the CPU utilization~\cite{fan2007power},
\begin{equation}
P(x_t) = P(0\%) + (P(100\%) - P(0\%))(2x_t - x_t^{1.4})
\end{equation}
where $x_t$ denotes the CPU utilization of the server at time $t$, $P(0\%)$ denotes the power consumption of the server in the idle mode, and $P(100\%)$ denotes the power consumption of the server in full load. In general, the power consumption of the server during the sleep to active transition is higher than $P(0\%)$ \cite{fan2007power,meisner2009powernap}.

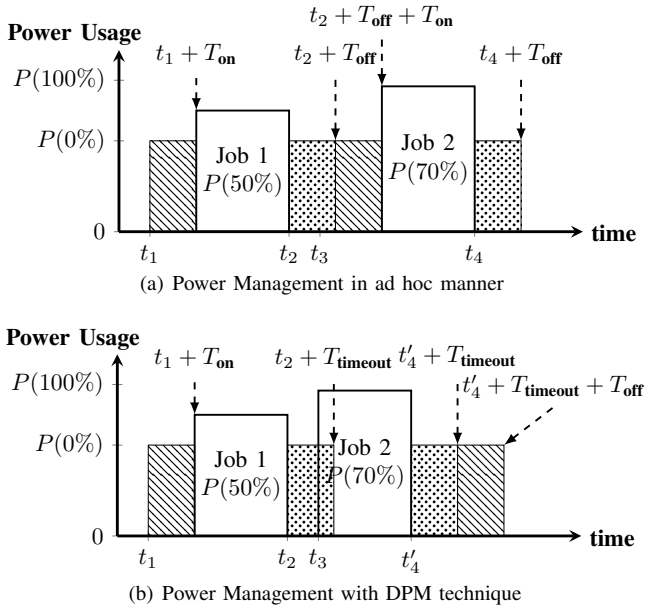
\begin{figure}
\centering
\subfigure[Power Management in ad hoc manner]{
\begin{tikzpicture}[scale=0.9]
	\begin{axis}[xmin=0,xmax=15,ymin=0,ymax=12,
			axis line style={very thick},
	        axis y line = left, ylabel=\bf{Power Usage},
	        ylabel near ticks,
	        y label style = {at={(axis cs:1,13)}, rotate=-90},
	        xlabel near ticks,
	        axis x line = bottom, xlabel=\bf{time},
	        x label style = {at={(axis cs:16, 1)}},
	        unit vector ratio=1 0.7 1,
	        xtick={1, 5.5, 6.5, 11.5},
	        xticklabels = {$t_1$, $t_2$, $t_3$,$t_4$},
	        ytick={0, 6, 10},
	        yticklabels = {$0$, $P(0\%)$, $P(100\%)$},
	        clip=false]
		\draw[thin, pattern=north west lines] (axis cs:1,0) rectangle (axis cs:2.5,6);
		\draw[thick] (axis cs:2.5,0) rectangle (axis cs:5.5,8) node[midway]
			{\begin{tabular}{c} Job 1\\$P(50\%)$\end{tabular}};
		\draw[thin, pattern=crosshatch dots] (axis cs:5.5,0) rectangle (axis cs:7,6);
		\draw[thin, pattern=north west lines] (axis cs:7,0) rectangle (axis cs:8.5,6);
		\draw[thick] (axis cs:8.5,0) rectangle (axis cs:11.5,9.6) node[midway]
			{\begin{tabular}{c} Job 2\\$P(70\%)$\end{tabular}};
		\draw[thin, pattern=crosshatch dots] (axis cs:11.5,0) rectangle (axis cs:13,6);
		\draw[dashed, latex-, thick] (axis cs:2.5,8) -- (axis cs:2.5,10.5) node[above] {$t_1 + T_\textbf{on}$};
		\draw[dashed, latex-, thick] (axis cs:7,6) -- (axis cs:7,10.5) node[above] {$t_2 + T_\textbf{off}$};
		\draw[dashed, latex-, thick] (axis cs:8.5,9.6) -- (axis cs:8.5,13) node[above]
			{$t_2 + T_\textbf{off} + T_\textbf{on}$};
		\draw[dashed, latex-, thick] (axis cs:13,6) -- (axis cs:13,10.5) node[above] {$t_4 + T_\textbf{off}$};
	\end{axis}
\end{tikzpicture}
}
\subfigure[Power Management with DPM technique]{
\begin{tikzpicture}[scale=0.9]
	\begin{axis}[xmin=0,xmax=15,ymin=0,ymax=12,
			axis line style={very thick},
	        axis y line = left, ylabel=\bf{Power Usage},
	        ylabel near ticks,
	        y label style = {at={(axis cs:1,13)}, rotate=-90},
	        xlabel near ticks,
	        axis x line = bottom, xlabel=\bf{time},
	        x label style = {at={(axis cs:16, 1)}},
	        unit vector ratio=1 0.7 1,
	        xtick={1, 5.5, 6.5, 9.5},
	        xticklabels = {$t_1$, $t_2$, $t_3$,$t_{4}'$},
	        ytick={0, 6, 10},
	        yticklabels = {$0$, $P(0\%)$, $P(100\%)$},
	        clip=false]
		\draw[thin, pattern=north west lines] (axis cs:1,0) rectangle (axis cs:2.5,6);
		\draw[thick] (axis cs:2.5,0) rectangle (axis cs:5.5,8) node[midway]
			{\begin{tabular}{c} Job 1\\$P(50\%)$\end{tabular}};
		\draw[thin, pattern=crosshatch dots] (axis cs:5.5,0) rectangle (axis cs:7,6);
		\draw[thick] (axis cs:6.5,0) rectangle (axis cs:9.5,9.6) node[midway]
			{\begin{tabular}{c} Job 2\\$P(70\%)$\end{tabular}};
		\draw[thin, pattern=crosshatch dots] (axis cs:9.5,0) rectangle (axis cs:11,6);
        \draw[thin, pattern=north west lines] (axis cs:11,0) rectangle (axis cs:12.5,6);
		\draw[dashed, latex-, thick] (axis cs:2.5,8) -- (axis cs:2.5,10.5) node[above] {$t_1 + T_\textbf{on}$};
		\draw[dashed, latex-, thick] (axis cs:7,6) -- (axis cs:7,10.5) node[above] {$t_2 + T_\textbf{timeout}$};
		\draw[dashed, latex-, thick] (axis cs:11,6) -- (axis cs:11,10.5) node[above] {$t_4' + T_\textbf{timeout}$};
        \draw[dashed, latex-, thick] (axis cs:12.5,6) -- (axis cs:14.1,8.5) node[above] {$t_4' + T_\textbf{timeout}+T_\textbf{off}$};
	\end{axis}
\end{tikzpicture}
}
\caption{Illustration of the effectiveness of server power management in the local tier.}
\label{Fig:power}
\end{figure}

We explain the effectiveness of power management in the local tier by Fig. \ref{Fig:power}. Assume that at time $0$ the server is in the sleep mode, and the arrival times and CPU resource utilizations of Jobs 1 and 2 are $<t_1$, $50\%>$ and $<t_3$, $70\%>$, respectively. Job $1$'s arrival triggers the server to switch to the active mode and start serving job $1$ from $t_1 + T_\textbf{on}$ to $t_2$. Fig. \ref{Fig:power} (a) illustrates the case when power management is performed in an ad hoc manner. In this case, the server turns back to the sleep mode from $t_2$, and the expected completion time of power mode switching is $t_2 + T_\textbf{off}$,
which is later than the arrival time of job $2$ ($t_3$). Thus the server switches back to active mode immediately after $t_2 + T_\textbf{off}$. 
At $t_2 + T_\textbf{off} + T_\textbf{on}$, the server starts serving job 2 and completes at time $t_4$. In this case, a portion of power consumption is wasted on the frequent turning ON/OFF of the server, and also extra latency is incurred on job 2. On the other hand, Fig. \ref{Fig:power} (b) illustrates the case when effective DPM technique is in place. In this case, an effective timeout is set after time $t_2$ and the server will stay in the idle state. If job 2 arrives before the timeout expires, the server will process job 2 immediately and finish at an earlier time $t'_4$, with $t'_4<t_4$. A simultaneous reduction in power/energy consumption and average job latency could be achieved through the effective DPM technique at the server level, and the DPM technique for each server could be performed in a distributed manner.

\begin{figure}
\centering
\includegraphics[width=0.8\columnwidth]{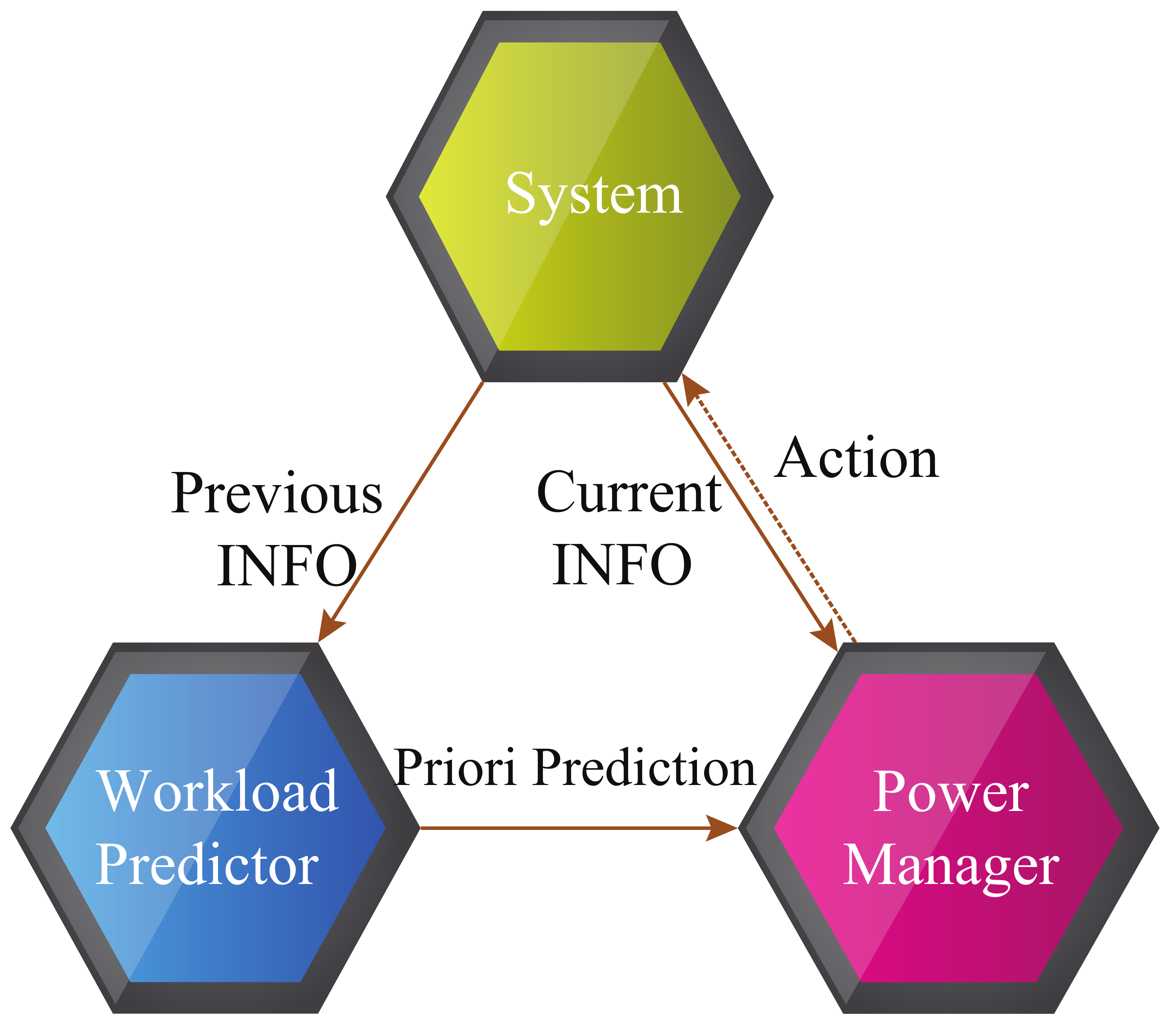}
\caption{Power management at each local server, comprising the workload predictor and the power manager.}
\label{Fig:powermanagement}
\end{figure}
The effective DPM technique in the local tier properly determines the most desirable timeout values in an online adaptive manner, based on the VM allocation results from the global tier. The DPM framework at the local server level is illustrated in Fig.~\ref{Fig:powermanagement}. A \emph{workload predictor} is incorporated in the DPM framework, which is critical to the performance of DPM by providing predictions of future workloads for the \emph{power manager}. The prediction together with the current information such as the number of pending jobs in the queue is fed into the power manager, and serve as the state of the power manager for selecting corresponding actions and learning in the observation domain of the system. 

The DPM framework of local servers relies heavily on a confident workload prediction and a properly designed power manager.
In \cite{wang2011deriving}, a Na\"\i ve Bayes classifier is adopted to perform the workload prediction. 
In this paper, we want to perform a more accurate (time-series) prediction with continuous values, and thus a \textit{recurrent neural network} (RNN) \cite{connor1994recurrent} or even a \textit{long short-term memory} (LSTM) network \cite{hochreiter1997long} becomes a good candidate for the workload predictor.
With accurate prediction and current information of the system under management, the power manager has to derive the most appropriate actions (timeout values) to help simultaneously reduce the power consumption of the server and the job latency, and the model-free RL technique \cite{tan2009adaptive,wang2011deriving} serves as a good candidate for the adaptive power management algorithm.

In Sections V and VI, we will describe the global and local tiers of the overall cloud resource allocation and power management framework. The global tier of cloud resource (VM) allocation exhibits high dimensions in state and action spaces, which prohibit the usefulness of traditional RL techniques. 
This is because the convergence speed of traditional RL techniques is in general proportional to the number of state-action pairs~\cite{sutton1998reinforcement}. Therefore, we adopt the emerging DRL technique~\cite{mnih2013playing}, which has the potential of handling large state space of complicated control problems, to solve the global tier problem. For the local tier of power management, we adopt a model-free RL-based DPM technique integrated with an effective workload predictor using the LSTM network.  

	\section{Overview of Deep Reinforcement Learning}
	\label{Sec:DRL}
    In this section, we present a generalized form of DRL technique compared with the prior work, which could be utilized for resource allocation and other problems as well.
The DRL technique is comprised of an offline DNN construction phase and an online deep Q-learning phase~\cite{mnih2013playing,mnih2015human}. The offline phase adopts DNN to derive the correlation between each state-action pair $(s,a)$ of the system under control and its value function $Q(s,a)$. The offline DNN construction phase needs to accumulate enough samples of $Q(s,a)$ value estimates and corresponding $(s,a)$ for constructing an enough accurate DNN, which can be a model-based procedure or from actual measurement data~\cite{Rao2009,silver2016mastering}. For the game playing applications ~\cite{mnih2013playing}, this procedure includes the pre-processing of game playing profiles/replays and obtaining the state transitioning profile and $Q(s,a)$ value estimates (e.g., win/lose or the score achieved). For the cloud resource allocation applications, we will make use of the real job arrival traces to obtain enough state transitioning profiles and $Q(s,a)$ value estimates, which can be a composite of power consumption, performance (job latency), and resiliency metrics, for the DNN construction. Arbitrary policy and gradually refined policy can be applied in this procedure. The transition profiles are stored in an experience memory $\mathcal{D}$ with capacity $N_\mathcal{D}$.
The use of experience memory can smooth out learning and avoid oscillations or divergence in the parameters.~\cite{mnih2013playing}. Based on the stored state transition profiles and $Q(s,a)$ value estimates, the DNN is constructed with weight set $\bf{\theta}$ trained using standard training algorithms~\cite{kingma2014adam}.

We summarize the key steps in the offline procedure, which are shown in line 1-4 in Algorithm~\ref{Alg:Sch}.

\begin{algorithm}
\caption{Illustration of the General DRL Framework}
\label{Alg:Sch}
\begin{algorithmic}[1]
\ENSURE
\STATE Extract real data profiles using certain control policies and obtain the corresponding state transition profiles and $Q(s,a)$ value estimates;
\STATE Store the state transition profiles and $Q(s,a)$ value estimates in experience memory $\mathcal{D}$ with capacity $N_\mathcal{D}$;
\STATE Iterations may be needed in the above procedure;
\STATE Pre-train a DNN with features $(s,a)$ and outcome $Q(s,a)$;

\REQUIRE
%
\FOR{each execution sequence}
	\FOR{at each decision epoch $t_k$}
		\STATE With probability $\epsilon$ select a random action, otherwise $a_k =\ argmax_{a} Q(s_k, a)$, in which $Q(s_k, a)$ is derived (estimated) from DNN;
		\STATE Perform system control using the chosen action;
		\STATE Observe state transition at next decision epoch $t_{k+1}$ with new state $s_{k+1}$, receive reward $r_k(s_k,a_k)$ during time period $[t_k,t_{k+1})$;
		\STATE Store transition $\left(s_k, a_k, r_k, s_{k + 1}\right)$ in $\mathcal{D}$;
		\STATE Updating $Q(s_k,a_k)$ based on $r_k(s_k,a_k)$ and $\max_{a'}Q(s_{k+1},a')$ based on Q-learning updating rule;
	\ENDFOR
\STATE Update DNN parameters $\theta$ using new Q-value estimates;
\ENDFOR
\end{algorithmic}
\end{algorithm}

The deep Q-learning technique is adopted for the online control based on the offline-trained DNN. More specifically, at each decision epoch $t_k$ of an execution sequence, the system under control is in a state $s_k$. The DRL agent performs inference using the DNN to derive the $Q(s_k,a)$ estimate of each state-action pair $(s_k,a)$, and uses $\epsilon$-greedy policy to derive the action with the highest $Q(s_k,a)$ with probability $1-\epsilon$ and choose the other actions randomly with total probability $\epsilon$. The chosen action is denoted by $a_k$. At the next decision epoch $t_{k+1}$, the DRL agent performs Q-value updating based on the total reward (or cost) $r_k(s_k,a_k)$ observed during this time period $[t_k,t_{k+1})$. At the end of the execution sequence, the DRL agent updates the DNN using the newly observed Q-value estimates, and the updated DNN will be utilized in the next execution sequence. More detailed procedures are shown in Algorithm~\ref{Alg:Sch}.

It can be observed from the above procedure that the DRL framework is highly scalable for large state space and could deal with the case of continuous state space, which is distinct from traditional RL techniques. On the other hand, the DRL framework requires a relatively low-dimensional action space because in each decision epoch the DRL agent needs to enumerate all possible actions at current state and perform inference using DNN to derive the optimal $Q(s,a)$ value estimate, which implies that the action space in the cloud resource allocation framework needs to be reduced.
    
	\section{The Global Tier of the Hierarchical Framework -- DRL-Based Cloud Resource Allocation}
	\label{Sec:globaltier}
    In this paper, we develop a scalable hierarchical framework for the overall cloud resource allocation and power management problem, comprising a global tier of cloud VM resource allocation and a local tier of power managements. The global tier adopts the emerging DRL technique to handle the high-dimensional state space in the VM resource allocation framework. In order to significantly reduce the action space, we adopt a continuous-time and event-driven decision framework in which each decision epoch coincides with the arrival time of a new VM (job) request. In this way the action at each decision epoch is simply the target server for VM allocation, which ensures that the available actions are enumerable at each epoch. The continuous-time Q-learning for SMDP~\cite{duff1995reinforcement} is chosen as the underlying RL technique in the DRL framework. 
    
    In the following we present the proposed DRL-based global tier of cloud resource allocation including novel aspects both in the offline and online phases.


\subsection{DRL-based Global Tier of Resource Allocation}
\label{Sec:drl-formation}
In the DRL-based global tier of cloud resource allocation, the job broker is controlled by the DRL agent,
and the server cluster is the environment. The DRL-based cloud resource allocation framework is continuous-time based and event-driven, in that the DRL agent selects an action at the time
of each VM (job) request arrival, i.e., a decision epoch. The state space, action space, and reward function of the DRL-based global tier of resource allocation are defined as follows:

\textbf{\emph{State Space:}} In the DRL-based resource allocation, we define the state at job $j$'s arrival time $t_j$, $s^{t_j}$,
as the union of the server cluster state at job $j$'s arrival time $s_c^{t_j}$ and the job $j$'s state $s_j$, i.e., $s^{t_j}=s_c^{t_j}\cup s_j$. All the $M$ servers can be equally divided into $K$ groups, $G_1, \cdots, G_K$.
We define the state of servers in group $G_k$ at time $t$ as $g_k^{t}$.
We also define the utilization requirement of resource type $p$ of job $j$ by $u_{jp}$, and the utilization level of server $m$ at time $t$ as $u_{mp}^{t}$. Therefore, the system state $s^{t_j}$ of the DRL-based cloud resource allocation tier can be represented using
$u_{jp}$'s and $u_{mp}^{t_j}$'s as follows:
\begin{eqnarray}
s^{t_j} & = & \left[ s_c^{t_j}, s_j\right] \nonumber =
				\left[ g_1^{t_j}, \cdots, g_K^{t_j}, s_j\right] \nonumber \\
		& = & [u_{11}^{t_j}, \cdots, u_{1|D|}^{t_j}, \cdots, u_{|M||D|}^{t_j},
			u_{j1}, \cdots, u_{j|D|}, d_j], \nonumber
\end{eqnarray}
where $d_j$ is the (estimated) job duration. The state space consists of all possible states and has a high dimension.

\textbf{\emph{Action space:}} The action of the DRL agent for cloud resource allocation is
defined as the index of server for VM (job) allocation. The action space
for a cluster with $M$ servers is defined as follows.
\begin{eqnarray}
\mathcal{A} = \left\{ a | a \in \{1, 2, \cdots, |M|\} \right\}  \nonumber
\end{eqnarray}
It can be observed that the action space is significantly reduced (to the same size as the total number of servers) by using an event-driven and continuous-time DRL-based decision framework.

\textbf{\emph{Reward:}} The overall profit of the server cluster equals to the total revenue of processing all the incoming jobs minus the total energy cost and the reliability penalty. The income achieved by processing a job decreases with the increase in job latency, including both waiting time in the queue and processing time in the server. \emph{Hot spot avoidance} is employed in physical servers because the overloading situations can easily lead to
resource shortage and affect hardware lifetime, and thereby undermining data center reliability. Similarly, for the sake of reliability, a cloud provider can introduce \emph{anti co-location} objectives to ensure spatial distances between VMs and the use of disjoint routing paths in the data center, so as to prevent a single failure from affecting multiple VMs belonging to the same cloud customer. It is clear that both load balancing and anti-colocation objectives are partially in conflict with power saving (and maybe job latency), as they actually try to avoid the usage of high VM consolidation ratios.
Through a joint consideration of power consumption, job latency, and reliability issues, we define the reward function $r(t)$ that the agent of the global tier receives as follows:
\begin{align}
r(t) = &-w_1\cdot Total\_Power(t)\\
&-w_2\cdot Number\_VMs(t)-w_3\cdot Reli\_Obj(t),\nonumber
\end{align}
where the three terms are the negatively weighted values of instantaneous total power consumption, number of VMs in the system, and reliability objective function value, respectively. Please note that according to the Little's Theorem \cite{gross2008fundamentals}, the average number of VMs pending in the system is proportional to the average VM (job) latency. Therefore, the DRL agent of the global tier optimizes a linear combination of total power consumption, VM latency, and reliability metrics when using the above instantaneous reward function.

\emph{Offline DNN Construction:} The DRL-based global tier of cloud resource allocation comprises an offline DNN construction phase and an online deep Q-learning phase. The offline DNN construction phase derives the correlation between Q value estimates with each state-action pair $(s^{t_j},a)$. A straightforward approach is to use a conventional feed-forward neural network to directly output Q value estimates. This works well with problems with relatively small state and action spaces (e.g., in~\cite{mnih2013playing}, the number of actions ranges from 2 to 18). Alternatively, we can train multiple neural networks in which each of them estimates the Q value for a subset of actions, for example, the Q values for assigning the job $j$ to one subset of servers $G_k$. However, this procedure will significantly increase the training time of the neural network, by up to a factor of $K$, compared with the single network solution. Moreover, this simple multiple neural networks method cannot well stress the importance of the related servers of a target action, which slows down the training process.

\begin{figure}
\centering
\resizebox{\columnwidth}{!}{
\begin{tikzpicture}[xscale=1, yscale=.85,
					nnblock/.style 2 args={
					  trapezium, trapezium angle=55, draw,inner xsep=0pt,outer sep=0pt,
					  minimum height=23, text width=10.5, label=center:#1, fill=#2,
					  opacity=.25, text opacity=1},
					shareweight/.style={dashed, arrows={angle 90[scale=1.1]-angle 90}, thick}]
					
	\node (in-state) at (7.5, 3.2) {$s^{t_j} = \left[g_1^{t_j} g_2^{t_j} s_j\right]$};
	\node (out-q) at (7.5, 9.) {Q value estimates for all actions};
					
	\node (ae1) [nnblock={\begin{tabular}{c}Auto-\\$\text{encoder}_1$\end{tabular}}{red}]
		at (5, 5) {};
	\node (ae2) [nnblock={\begin{tabular}{c}Auto-\\$\text{encoder}_2$\end{tabular}}{red}]
		at (10, 5) {};
	\draw[shareweight] (ae1.east) -- (ae2.west) node[midway, fill=white, inner sep=-2]
		{\begin{tabular}{c} Share\\Weight\end{tabular}};
		
	\node (q1) [nnblock={$\text{Sub-Q}_1$}{blue}] at (5, 7.4) {};
	\node (q2) [nnblock={$\text{Sub-Q}_2$}{blue}] at (10, 7.4) {};
	\draw[shareweight] (q1.east) -- (q2.west) node[midway, fill=white, inner sep=-2]
		{\begin{tabular}{c} Share\\Weight\end{tabular}};
		
	\draw[-{Latex[scale=1.2]}] ($(in-state.north) + (0., -0.1)$) -- ++(0, 0.4) -| (ae1.south);
	\draw[-{Latex[scale=1.2]}] (in-state.north) -- ++(0, 0.3) -- ++(-4.1, 0) -- ++(0, 2.1) -| ($(q1.south) + (-0.5, 0)$);
	\node[circle,fill,inner sep=1.2] at ($(ae1.south) + (0, -0.5)$) {};
	
	\draw[-{Latex[scale=1.2]}] ($(in-state.north) + (0.4, -0.1)$) -- ++(0, 0.4) -| (ae2.south);
	\draw[-{Latex[scale=1.2]}] ($(in-state.north) + (0.4, -0.1)$) -- ++(0, 0.4) -- ++(3.7, 0) -- ++(0, 2.1) -| ($(q2.south) + (-0.5, 0)$);
	\node[circle,fill,inner sep=1.2] at ($(ae2.south) + (0, -0.5)$) {};
	
	\draw[-{Latex[scale=1.2]}] ($(in-state.north) + (0.8, -0.1)$) -- ++(0, 0.15) -- ++(3.8, 0)
								 -- ++(0, 2.6) -| ($(q2.south) + (0.5, 0)$);
	\draw[-{Latex[scale=1.2]}] ($(in-state.north) + (0.8, -0.1)$) -- ++(0, 0.15) -- ++(3.8, 0)
								 -- ++(0, 2.6) -| ($(q1.south) + (0.5, 0)$);
	\node[circle,fill,inner sep=1.2] at ($(q2.south) + (0.5, -0.55)$) {};
								
	\draw[-{Latex[scale=1.2]}] (ae1.north) -- ++(0, 0.2) -| (q2.south);
	\draw[-{Latex[scale=1.2]}] ($(ae2.north) + (-0.5, 0)$) -- ++(0, 0.45) -| (q1.south);
	
	\draw[-{Latex[scale=1.2]}] (q1.north) -- ++(0, 0.3) -| ($(out-q.south) + (-0.85, .1)$);
	\draw[-{Latex[scale=1.2]}] (q2.north) -- ++(0, 0.3) -| ($(out-q.south) + (0.85, .1)$);
\end{tikzpicture}
}
\vskip -1em
\caption{Deep neural network for Q-value estimating with autoencoder and weight sharing scheme ($K$ = 2).}
\label{Fig:nn}
\end{figure}
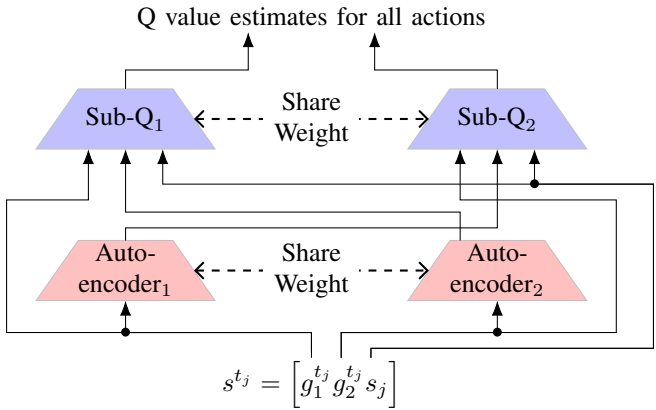

In order to address these issues, we harness the power of \emph{representation learning} and \emph{weight sharing} for DNN construction, with basic procedure shown in Fig.~\ref{Fig:nn}. Specifically, we first use an autoencoder to extract a lower-dimensional high-level representation of server group state $g_k^{t_j}$ for each possible $k$ value, denoted by $\bar{g}_k^{t_j}$. Next, for estimating the Q value of the action of allocating VM to servers in $G_k$,
the neural network $\textbf{Sub-Q}_k$ takes $g_k^{t_j}$, $s_j$ and all $\bar{g}_{k^\prime}^{t_j}\ (k^\prime \neq k)$
as input features. The dimension difference between $g_k^{t_j}$ and $\bar{g}_{k^\prime}^{t_j}\ (k^\prime \neq k)$ reflects the importance of the targeting server group's own state compared with the other server groups, which determines the degree of reduction in the state space.

In addition, we introduce weight sharing among all $K$ autoencoders, as well as all $\textbf{Sub-Q}_k$'s. Weight sharing has been successfully applied to convolutional neural networks and image processing~\cite{lecun1995convolutional}. It is used in our model due to following reasons: 1) With weight sharing, any training samples can be used to train the $\textbf{Sub-Q}_k$'s and autoencoders,
compared to the case without weight sharing, where only the training samples allocated to a server in $G_k$ can be used to train $\textbf{Sub-Q}_k$. This usually
leads to higher degree of scalability. 2) Weight sharing reduces the total
number of required parameters and the training time.

\emph{Online Deep Q-Learning:} The online deep Q-learning phase, which is an integration of the DRL discussed in Section III and the Q-learning for SMDP, is utilized for action selection and value function/DNN updating. At each decision epoch $t_j$, it performs inference using the DNN (with autoencoder incorporated) to derive the $Q(s^{t_j},a)$ value estimate of each state-action pair $(s^{t_j},a)$, and use $\epsilon$-greedy policy for action selection. At the next decision epoch $t_{j+1}$, it performs value updating using Eqn. (2) in Q-learning for SMDP. At the end of each execution sequence, it updates the DNN with autoencoders based on the updated Q value estimates.

\subsection{Convergence and Computational Complexity Analysis of the Global Tier}
\label{Sec:drl-converge}
It has been proven that the Q-learning technique will gradually converge to the optimal policy under stationary Markov decision process (MDP) environment and sufficiently small learning rate~\cite{watkins1992q}. Hence, the proposed DRL-based resource allocation framework will converge to the optimal policy when (i) the environment evolves as a stationary, memoryless SMDP and (ii) the DNN is sufficiently accurate to return the action associated with the optimal $Q(s,a)$ estimate. In reality, the stationary property is difficult to satisfy as actual cloud workloads are time-variant. However, simulation results will demonstrate the effectiveness of the DRL-based global tier in realistic, non-stationary cloud environments.

The proposed DRL-based global tier of resource allocation exhibits low online computational complexity, i.e., the computational complexity is proportional to the number of actions (number of servers) at each decision epoch (upon arrival of each VM), which is insignificant for cloud computing systems.
	\section{The Local Tier of the Hierarchical Framework -- RL-Based Power Management for Servers}
    \label{Sec:localtier}
In this section, we describe the proposed local tier of power management in the hierarchical framework, which is responsible for controlling the turning ON/OFF of local servers in order to simultaneously reduce the power consumption and the average job latency. The local tier includes the workload predictor using the LSTM network and the adaptive power manager based on the model-free, continuous-time Q-learning for SMDP. Details will be described next.
\subsection{Workload Predictor Using the Long Short-Term Memory (LSTM) Network}
The workload predictor is responsible for providing partial observation of the actual future workload characteristics (i.e., inter-arrival times of jobs) to the power manager, which will be utilized for defining system states in the RL algorithm. The workload characteristics of servers are in fact the result of the global tier of VM resource allocation. Previous works on workload prediction in \cite{Srivastava:1996:PSS:231913.231922,643266} assume that a linear combination of previous idle times (or request inter-arrival times) may be used to infer the future ones, which is not always true. For example, one very long inter-arrival time can ruin a set of subsequent predictions. In order to overcome the above effect and achieve much higher prediction accuracy, in this work we adopt the LSTM neural network \cite{hochreiter1997long} for workload prediction, which is a recurrent neural network architecture and can be applied for prediction of time series sequences. The LSTM network can eliminate the vanishing gradient problem and capture the long-term dependencies in the time series, and works efficiently where \emph{back propagation through time} (BPTT) technique is adopted \cite{kalchbrenner2015grid}.

\begin{figure}
		\centering
		\includegraphics[width=\columnwidth]{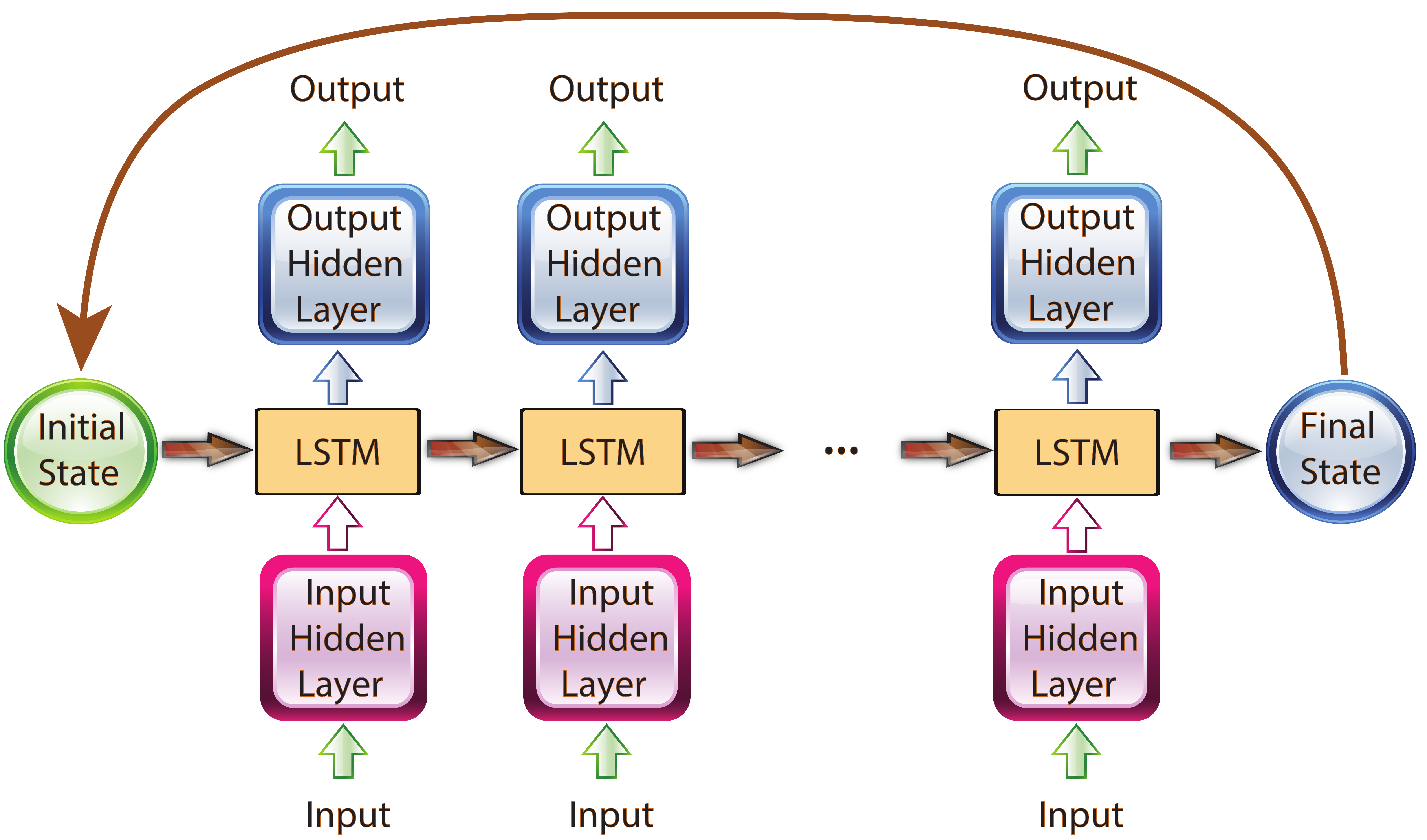}
		\vskip -0.8em
		\caption{Unrolled LSTM neural network for workload prediction at the local server.}
		\label{fig:LSTM}	
	\end{figure}

 The LSTM network structure is shown in Fig. \ref{fig:LSTM}. The network has three layers, input hidden layer, LSTM cell layer and output hidden layer. In the proposed LSTM network for workload prediction, we predict the next job inter-arrival time based on the past 35 inter-arrival times as the look-back time steps. The size of both input and output of an LSTM cell should be 1 according to the dimension of the measured trace.
We set 30 hidden units in each LSTM cell, and all LSTM cells have shared weights. In our LSTM-based workload prediction model, we discretize the output inter-arrival time prediction by setting $n$ predefined categories, corresponding to $n$ different states in the RL algorithm utilized in the power manager.

In the training process, first we initialize the weights for the input layer and output layer as a normal distribution with a mean value of 0 and standard deviation of 1. The bias for both layers is set as a constant value 0.1. The initial state of LSTM cell is set as 0 for all cells. In response to the back propagated errors, the network is updated by adopting \emph{Adam optimization} \cite{kingma2014adam}, a method for efficient stochastic optimization that only requires first-order gradients with little memory requirement. 
The method computes individual adaptive learning rates from estimates of the first and second moments of the gradients \cite{DBLP:journals/corr/KingmaB14}. The state of the LSTM cell and weights will be trained for minimizing the propagated errors.

\subsection{Distributed Dynamic Power Management for Local Servers}
\label{Sec:DPM}
In this subsection, we describe the proposed adaptive power manager for local servers using model-free RL, based on effective workload prediction results. The continuous-time Q-learning for SMDP is adopted as the RL technique in order to reduce the decision frequency and overheads. The power management for local servers is performed in a distributed manner. First, we provide a number of key definitions and notations.
Let $t^{G}_{j}\ (j=1,2,3...)$ denote the time when the $j$-th job (VM) arrives in the local machine. Suppose that we are currently at time $t$ with $t^{G}_{j} \leq t < t^{G}_{j+1}$. Note that the power manager has no exact information on $t^{G}_{j+1}$ at the current time $t$. Instead, it only receives an
estimation of inter-arrival time from the LSTM-based workload predictor.

The power manager monitors the following \emph{state parameters} of the power managed system (server):
\begin{enumerate}
\item  The machine power state $M(t)$, which can be active, idle, sleep, etc.
\item  The estimated next job inter-arrival time from the workload predictor.
\end{enumerate}

To apply RL techniques to DPM frameworks, we define
\emph{decision epochs}, i.e., when new decisions are made and updates for the RL algorithm are executed. 
In our case, the decision epochs coincide with one of the following three cases:
\begin{enumerate}
\item The machine enters the idle state (usage = 0) and $JQ(t) = 0$ (no waiting job in the buffer queue).
\item The machine is in the idle state and a new job arrives.
\item The machine is in the sleep state and a new job arrives.
\end{enumerate}

The proposed RL-based DPM operates as follows. At
each decision epoch, the power manager finds itself in one of the three
aforesaid conditions and it will make a decision. If it finds itself in case (1),
it will use the RL-based timeout policy. A list of timeout periods, including the immediate shutdown (timeout value = 0), serve as the action set $\mathcal{A}$ in this case, and the power manager learns to choose the optimal action by using RL technique. If it
is in case (2), it will start executing the new job and turn from idle to active. If it is in case (3), the server will turn active from the sleep state and start the new job as soon as it is ready.
Updates are no need to perform in cases (2) and (3) because there is just one action to choose. As pointed out in reference \cite{931003}, the optimal policy when the machine is idle
for non-Markov environments is often a timeout policy, wherein
the machine is turned off to sleep if it is idle for more than a specified timeout
period.
 
In this local RL-based DPM framework, we use the reward rate defined as follows:
\begin{equation}
r(t)= -wP(t)-(1-w)JQ(t)
\end{equation}
The reward rate function is the negative of a linearly-weighted combination of power consumption $P(t)$ of server and the number of jobs $JQ(t)$ buffered in the service queue, where $w$ is the weighting factor.
This is a reasonable reward rate because as authors in~\cite{qiu2007stochastic} has pointed out, the average number of buffered service requests/jobs is
proportional to the average latency for each job, which
is defined as the average time for each job between the
moment it is generated and the moment that the server finishes
processing it, i.e., it includes the queueing time plus execution
time. The above observation is based on the Little's
Law \cite{little2008little}. In this way, the value function $Q(s,a)$ for each state-action pair $(s,a)$ is a combination of the expected total
discounted energy consumption and total latency experienced
by all jobs. Since the total number of jobs and the total execution time are fixed,
the value function is equivalent to a linear combination of
the average power consumption and average per-job latency. The relative weight $w$ between the average power consumption and latency can be adjusted to obtain the power-latency trade-off curve. The detailed procedure of the RL algorithm is in Algorithm \ref{Alg:RL-based DPM}.

\begin{algorithm}
\caption{The RL-based DPM framework in the local tier.}
\label{Alg:RL-based DPM}
\begin{algorithmic}[1]
\STATE At each decision epoch $t^{D}$, denote the RL state as $s(t^{D})$ for the power-managed system.
\FOR{each decision epoch $t_k^{D}$}
	\STATE With probability $\epsilon$ select a random action from action set $\mathcal{A}$ (timeout values), otherwise $a_k =\ argmax_{a} Q(s(t_k^{D}), a)$.
	\STATE Apply the chosen timeout value $a_k$. 
    \STATE If job arrives during the timeout period, turn active to process the job until the job queue is empty. Otherwise turn sleep until the next job arrives.
	\STATE Then we arrive at the next
decision epoch $t^{D}_{k+1}$.
	\STATE Observe state transition at next decision epoch $t^{D}_{k+1}$ with new state $s(t^{D}_{k+1})$, and calculate reward rate during time period $[t^{D}_{k},t^{D}_{k+1})$.
	\STATE Updating $Q(s(t^{D}_{k}),a_k)$ based on reward rate and $\max_{a'}Q(s(t^{D}_{k+1}),a')$ based on the updating rule of Q-learning for SMDP (Eqn. (2)).
\ENDFOR
\end{algorithmic}
\end{algorithm}
	\section{Experimental Results}
	\label{Sec:experiment}
    In this section, we first describe the simulation setup.
    Then, from perspectives of power consumption and job latency we compare our proposed hierarchical framework for resource allocation and power management with the DRL-based framework for resource allocation ONLY, and the baseline round-robin VM allocation policy. Finally, we evaluate our proposed framework by investigating the optimal trade-off between the power consumption and latency.
\subsection{Simulation Setups}
Without loss of generality, in this paper, we assume a homogeneous server cluster.
The peak power of each server is $P(100\%) = 145$W, and the idle power consumption is $P(0\%) = 87$W~\cite{fan2007power}. We set the server power mode transition times $T_\textbf{on} = 30$s and $T_\textbf{off} = 30$s. The number of machines in each cluster $M$ is set as $30$, $40$ respectively. Please note that our proposed framework is applicable to the case with more servers as well.

We use real data center workload traces from Google cluster-usage traces~\cite{clusterdata:Reiss2011}, which provide the server cluster usage data over a month-long period in May 2011. The extracted job traces include job arrival time (absolute time value), job duration, and resource requests of each job, which include CPU, memory and disk requirements (normalized by the resource of one server). All the extracted jobs are with a duration between 1 minute and 2 hours, ordered increasingly based on their arrival time.
To simulate the workload on a 30-40 machines cluster, we split the traces into 200 segments, and each segment contains about 100,000 jobs, corresponding to the workload for a $M$-machine cluster in a week.

In this work, for the global tier, we perform offline training, including the experience memory initialization and training of the autoencoder, using the whole Google cluster traces. To obtain DNN model in the global tier of the proposed framework, we use workload traces for five different $M$-machine clusters. We generate four new state transition profiles using the $\epsilon$-greedy policy \cite{sutton1996generalization}
and store the transition profiles in the memory before sampling the minibatch for training the DNN. In addition, we clip the gradients to make their norm values less than or equal to 10. The gradient clipping method has been introduced to DRL in~\cite{wang2015dueling}.

In DNN construction, we use two layers of fully-connected Exponential Linear Units (ELUs) with 30 and 15 neurons, respectively, to build an autoencoder. Each $\textbf{Sub-Q}_k$ mentioned in Section \ref{Sec:globaltier} contains a single fully-connected hidden layer with 128 ELUs, and a fully-connected linear layer with a single output for each valid action in the group. The number of groups varies between 2 and 4. We use $\beta = 0.5$ in simulations for Q-learning discount rate.

\subsection{Comparison with Baselines}

We simulate five different one-week job traces using the proposed hierarchical framework with different parameters and compare the performance against the DRL-based resource allocation framework and the round-robin allocation policy (denoted as the baseline) in terms of the (accumulative) power consumption and the accumulated job latency.
\begin{figure}
	\centering
    \subfigure[Accumulated job latency versus the number of jobs]{
	\includegraphics[width=\columnwidth]{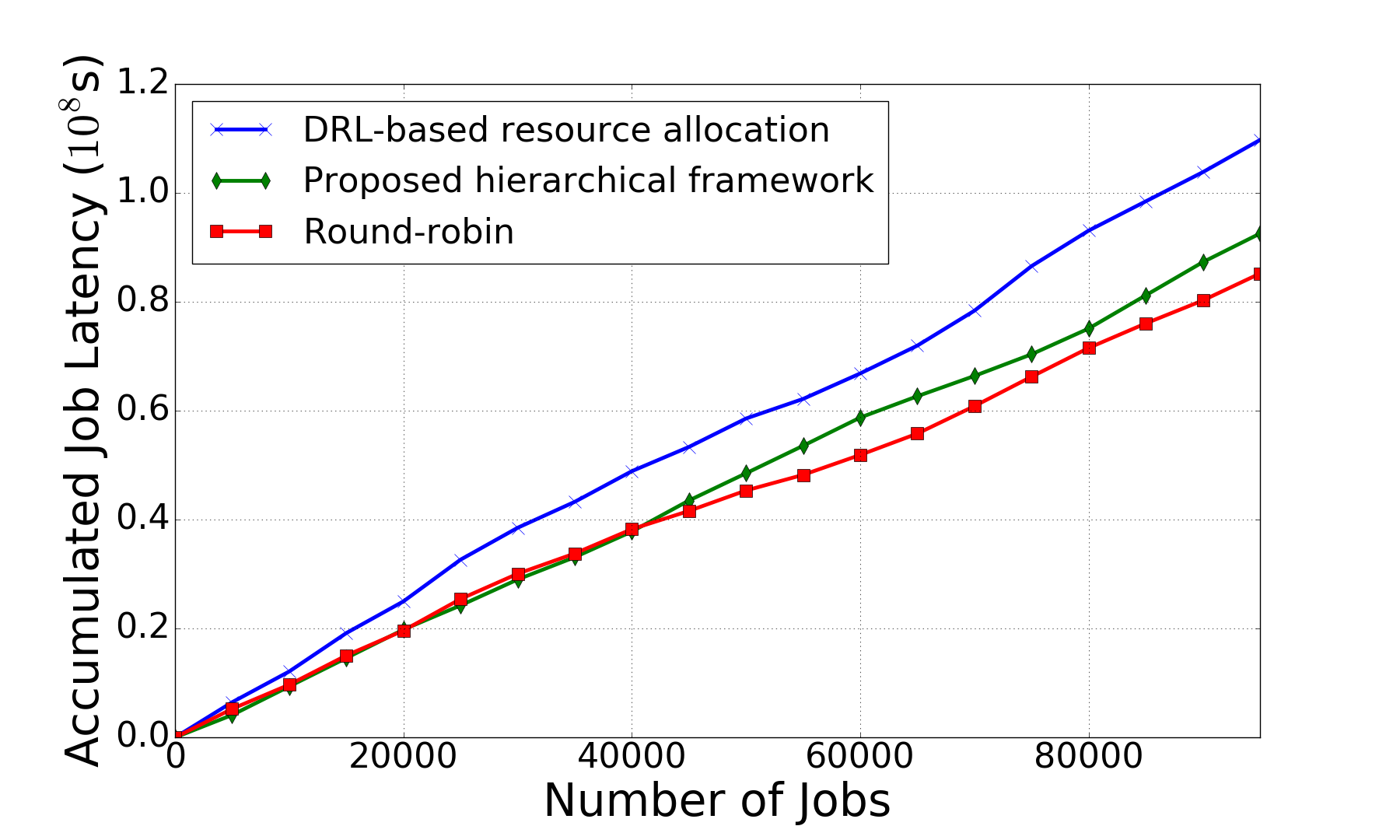}
	\label{Fig:30-machine-latency}
	}
    \vskip -0.5em
	\subfigure[Energy usage versus the number of jobs]{
	\includegraphics[width=\columnwidth]{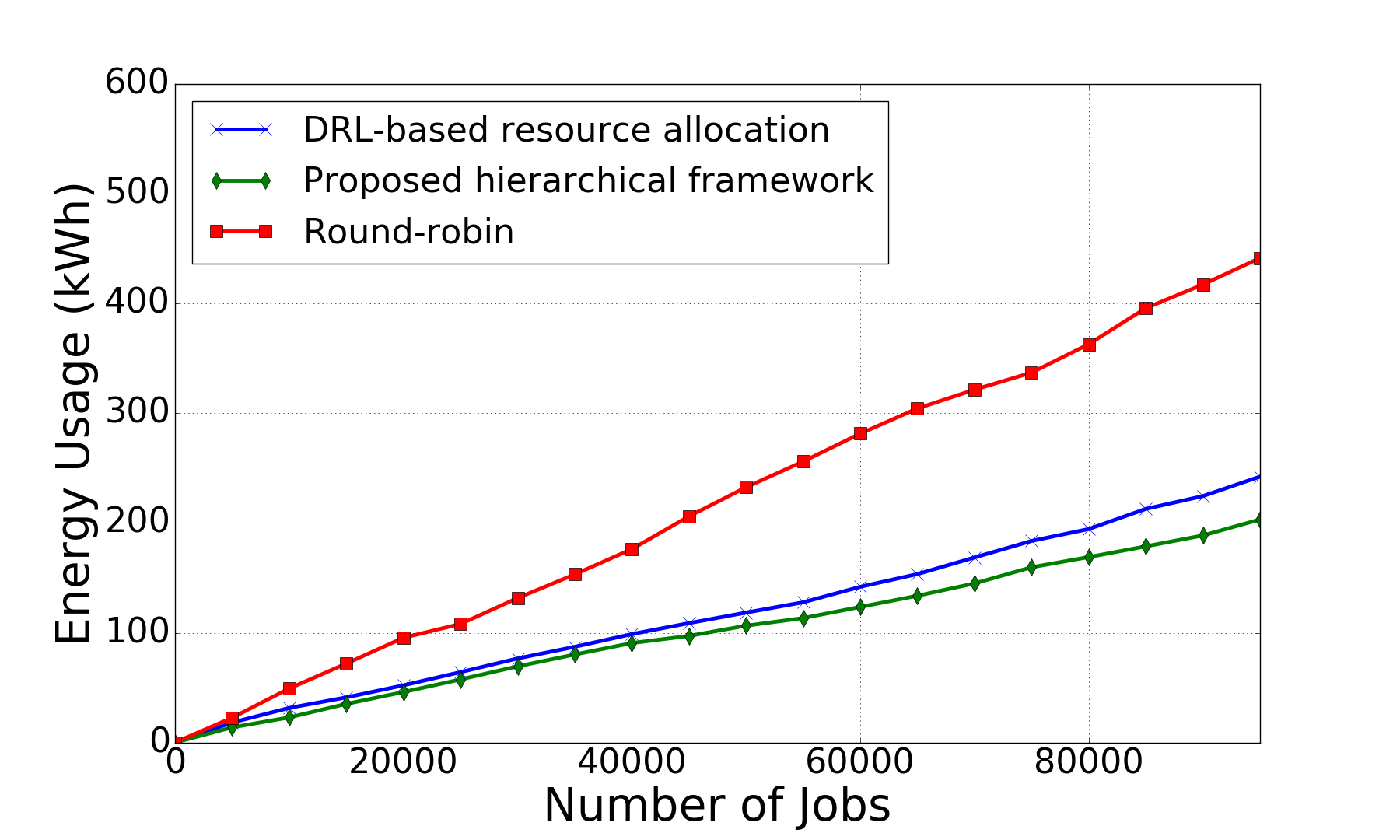}
	\label{Fig:30-power}
	}
	\vskip -0.8em
	\caption{Comparison among the proposed hierarchical framework, the DRL-based resource allocation framework, and the round-robin baseline in the case of $M = 30$.}
	\label{Fig:30-machine}
\end{figure}

Fig.~\ref{Fig:30-machine} shows the experimental results when the number of machines in the cluster $M = 30$.
Compared with the round-robin method, the hierarchical framework and DRL-based resource allocation result in longer latency than the baseline round-robin method shown in Fig.~\ref{Fig:30-machine-latency}. This is because in the round-robin method jobs are dispatched evenly to each machine, and generally, the jobs do not need to wait in each machine job queue.
However, from the perspective of energy (accumulative power consumption) usage shown in Fig.~\ref{Fig:30-power}, the round robin method gives a larger increase rate than the hierarchical framework or DRL-based resource allocation, which implies the round robin method has a larger power consumption.
\begin{table}[b]
\centering
\caption{Summary of the server cluster performance metrics (accumulated energy, accumulated latency, and average power consumption) with job number = 95000}\label{Table1}
\vskip -0.8em
\resizebox{\columnwidth}{!}{
\begin{tabular}{c|c|c|c}
\hline
 Round-Robin& Energy (kWh) & Latency($10^{6}$s) & Power (W)\\
\hline
$M = 30$ & 441.47 & 85.20 & 2627.79\\
\hline
$M = 40$ & 561.13 & 85.20 & 3340.06\\
\hline
DRL-based& Energy (kWh) & Latency($10^{6}$s)& Power (W) \\
\hline
$M = 30$ & 242.25 & 109.73  & 1441.96\\
\hline
$M = 40$ & 273.41 & 108.76 & 1627.44\\
\hline
Hierarchical& Energy (kWh) & Latency($10^{6}$s) & Power (W)\\
\hline
$M = 30$ & 203.21 & 92.53  & 1209.58\\
\hline
$M = 40$ & 224.51 & 94.26 & 1336.37\\
\hline
\end{tabular}
}
\end{table}
On the other hand, compared with the DRL-based resource allocation, the proposed hierarchical framework shows a reduced job latency shown in Fig.~\ref{Fig:30-machine-latency} as well as a lower energy usage shown in Fig.~\ref{Fig:30-power}. From Fig.~\ref{Fig:30-power}, we can also observe the energy chart for the proposed framework is always lower than that for DRL-based resource allocation or round robin baseline, which means our proposed hierarchical framework achieves the lowest power consumption among three. Shown in Table~\ref{Table1}, given the number of jobs of $95,000$, compared with the round-robin method, our proposed hierarchical framework saves 53.97\% power and energy. It also saves 16.12\% power/energy and 16.67\% latency compared with the DRL-based resource allocation framework. 

\begin{figure}
	\centering
    \subfigure[Accumulated job latency versus the number of jobs]{
	\includegraphics[width=\columnwidth]{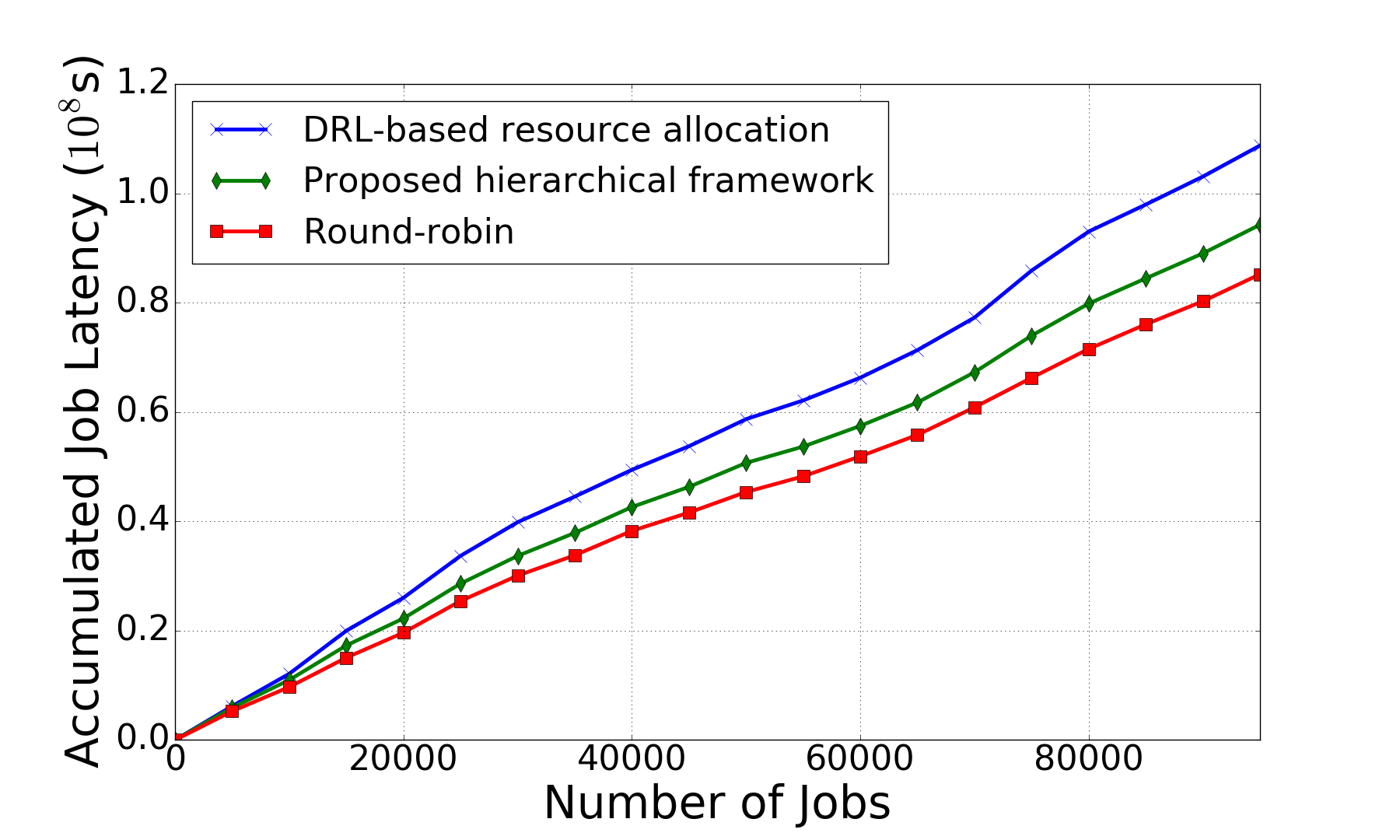}
	\label{Fig:40-machine-latency}
	}
    \vskip -0.5em
	\subfigure[Energy usage versus the number of jobs]{
	\includegraphics[width=\columnwidth]{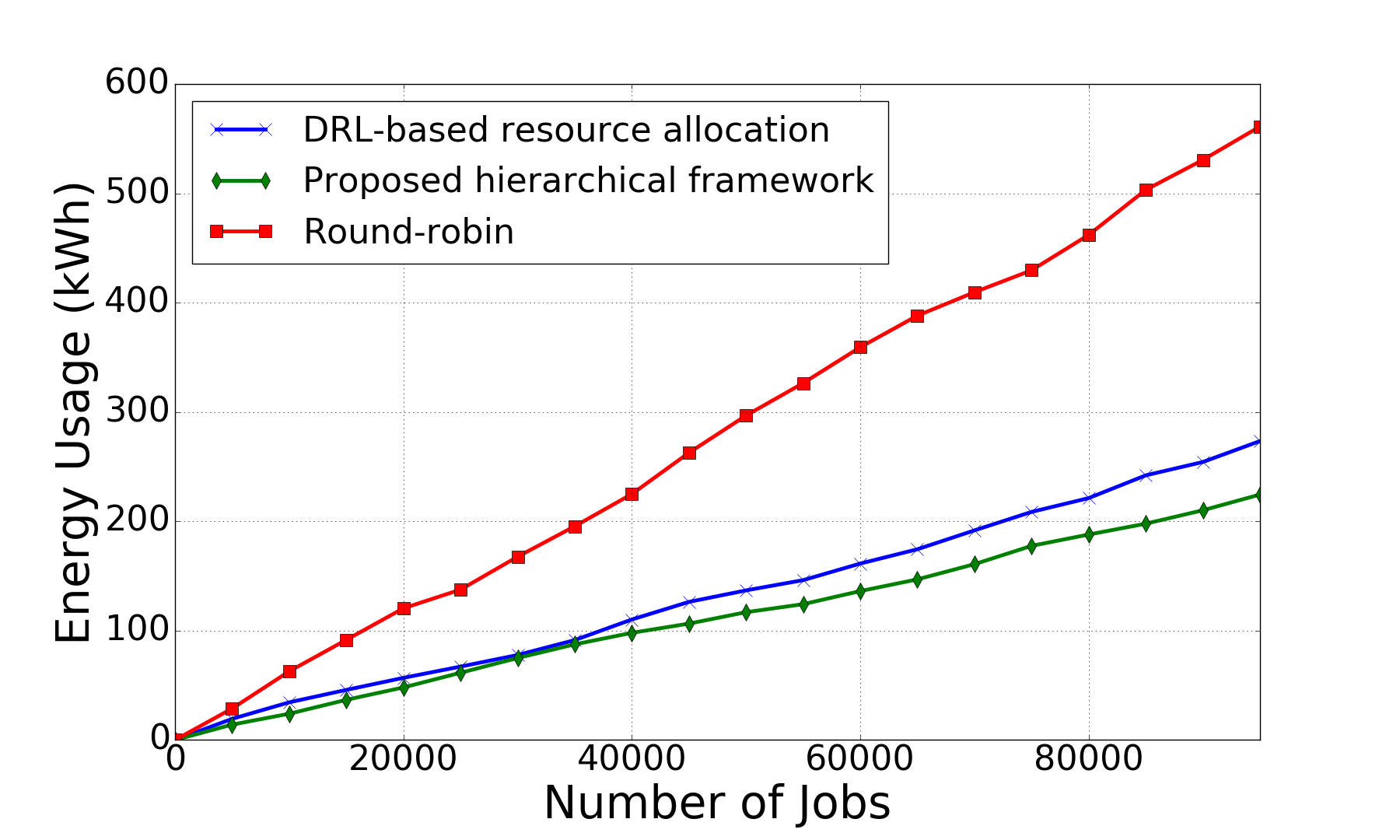}
	\label{Fig:40-power}
	}
	\vskip -0.8em
	\caption{Comparison among the proposed hierarchical framework, the DRL-based resource allocation framework, and the round-robin baseline in the case of $M = 40$.}
	\label{Fig:40-machine}
\end{figure}

Similarly, in the case of 40 machines in the cluster as shown in Table~\ref{Table1}, given the number of jobs of $95,000$, the proposed hierarchical framework consumes 59.99\% less power and energy compared with round-robin baseline, as well as 17.89\% less power and energy compared with the DRL-based resource allocation framework. The latency of the hierarchical framework reduces by 13.32\% compared with the DRL-based resource allocation framework. From Fig.~\ref{Fig:30-machine-latency},~\ref{Fig:40-machine-latency}, we can observe that the corresponding hierarchical frameworks' latency increase rates in two figures differ very little and so as to the corresponding DRL-based resource allocation frameworks. In Fig.~\ref{Fig:30-power} and Fig.~\ref{Fig:40-power}, the trend of energy usage of corresponding hierarchical frameworks and DRL-based resource allocation frameworks remains close which means their power consumption remains as the number of machines increases. However, the increase rate of energy usage (power) for round-robin methods becomes larger as the number of machines $M$ increases. Thus, when the number of jobs increases, in a larger-size server cluster, the round robin baseline system consumes an unacceptable amount of power/energy.
These facts imply that the DRL-based resource allocation framework is capable of managing a large-size server cluster with an enormous number of jobs. 
With the help of local and distributed power manager, the proposed hierarchical framework achieves shorter latency and less power/energy consumption.


\subsection{Trade-off of Power/Energy Consumption and Average Latency}
	 \begin{figure}
		\centering
		\includegraphics[width=\columnwidth]{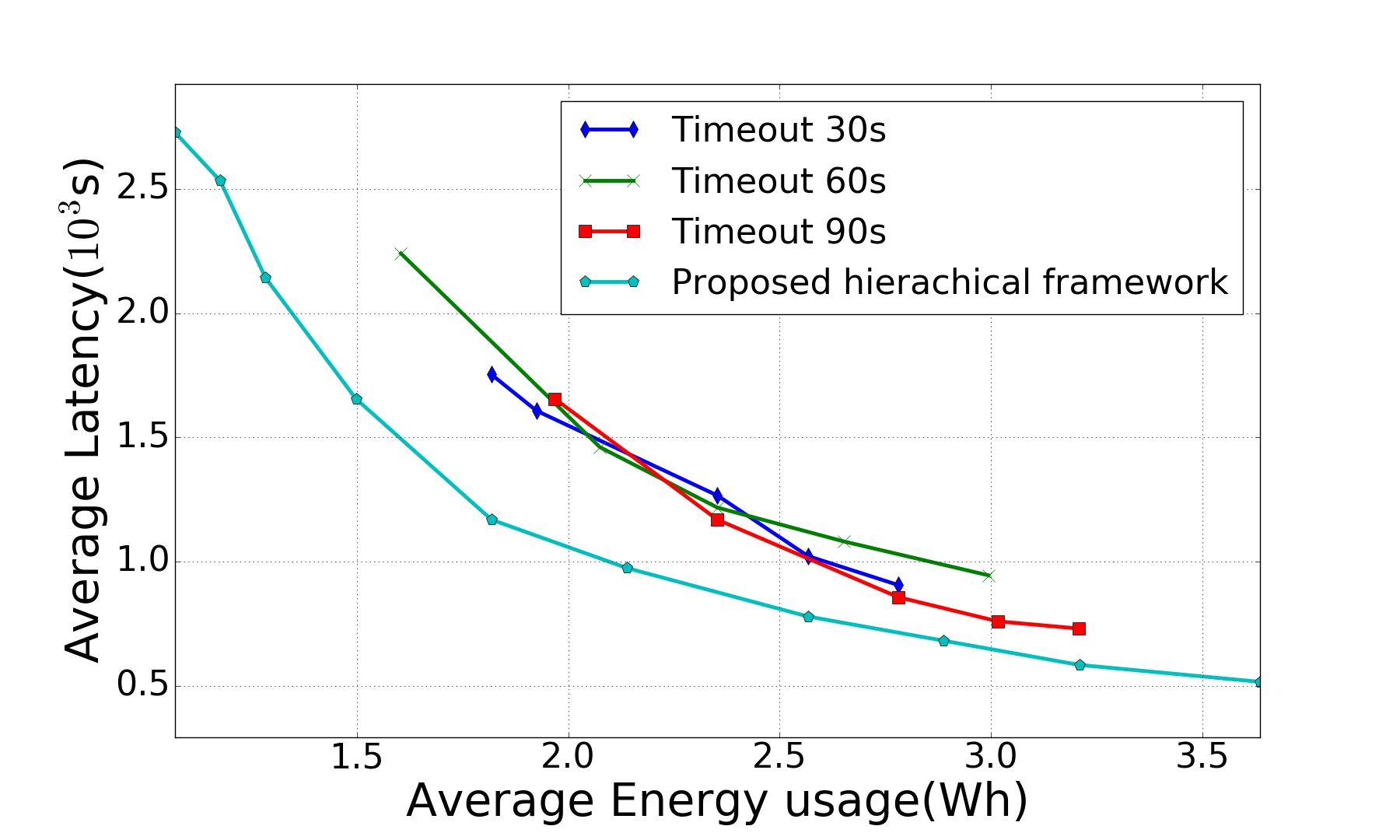}
		\vskip -0.8em
		\caption{Trade-off curves between average latency and average energy consumption of the proposed hierarchical framework and baseline systems.}
		\label{fig:tradeoff}
	\end{figure}	
Next, we explore the power (energy) and latency trade-off curves for the proposed framework as shown in Fig.~\ref{fig:tradeoff}. The latency and energy usage are averaged for each job. We first set three baselines with the DRL-based resource allocation tier and the local tier with different fixed timeout values, which is discussed in Section \ref{Sec:DPM}. The timeout values of the fixed timeout baselines are set to be $30s$, $60s$, and $90s$, respectively. Please note when the timeout value is fixed, the baseline system cannot reach every point in the energy and latency space, so that the curves for baselines are not complete. We can observe the proposed hierarchical framework can achieve the smallest area against the axes of power and latency, which denotes that our proposed hierarchical framework gives the best trade-off than any of those with a fixed timeout value.
For instance, compared with a baseline with fixed timeout value of 60, the maximum average latency saving with the same energy usage is 14.37\%, while the maximum power/energy saving with the same average latency is 16.13\%; compared with a baseline with fixed timeout value of 90, the maximum average latency saving with the same energy usage is 16.16\%, while the maximum average power/energy saving with the same latency is 16.20\%

	\section{Conclusion}
	\label{Sec:conclusion}
In this paper, a hierarchical framework is proposed to solve the resource allocation problem and power management problem in the cloud computing.
The proposed hierarchical framework comprises a global tier for VM resource allocation to the servers and a local tier for power management of local servers. Besides the enhanced scalability and reduced state/action space dimensions, the proposed hierarchical framework enables to perform the local power managements of servers in an online and distributed manner, which further enhances the parallelism degree and reduces the online computational complexity.
The emerging DRL technique is adopted to solve the global tier problem while an autoencoder and a novel weight sharing structure are adopted for acceleration.
For local tier, an LSTM based workload predictor helps a model-free RL based power manager to determine the suitable action of the servers.
Experiment results using actual Google cluster traces show that proposed hierarchical framework significantly save the power consumption/energy usage than the baseline while achieves similar average latency. 
Meanwhile, the proposed framework can achieve the best trade-off between latency and power/energy consumption in a server cluster. 

	\bibliographystyle{IEEEtran}

	\bibliography{reference}

\begin{thebibliography}{10}
\providecommand{\url}[1]{#1}
\csname url@samestyle\endcsname
\providecommand{\newblock}{\relax}
\providecommand{\bibinfo}[2]{#2}
\providecommand{\BIBentrySTDinterwordspacing}{\spaceskip=0pt\relax}
\providecommand{\BIBentryALTinterwordstretchfactor}{4}
\providecommand{\BIBentryALTinterwordspacing}{\spaceskip=\fontdimen2\font plus
\BIBentryALTinterwordstretchfactor\fontdimen3\font minus
  \fontdimen4\font\relax}
\providecommand{\BIBforeignlanguage}[2]{{%
\expandafter\ifx\csname l@#1\endcsname\relax
\typeout{** WARNING: IEEEtran.bst: No hyphenation pattern has been}%
\typeout{** loaded for the language `#1'. Using the pattern for}%
\typeout{** the default language instead.}%
\else
\language=\csname l@#1\endcsname
\fi
#2}}
\providecommand{\BIBdecl}{\relax}
\BIBdecl

\bibitem{mishra2010towards}
A.~K. Mishra, J.~L. Hellerstein, W.~Cirne, and C.~R. Das, ``Towards
  characterizing cloud backend workloads: insights from google compute
  clusters,'' \emph{ACM SIGMETRICS Performance Evaluation Review}, vol.~37,
  no.~4, pp. 34--41, 2010.

\bibitem{khan2012workload}
A.~Khan, X.~Yan, S.~Tao, and N.~Anerousis, ``Workload characterization and
  prediction in the cloud: A multiple time series approach,'' in \emph{2012
  IEEE Network Operations and Management Symposium}.\hskip 1em plus 0.5em minus
  0.4em\relax IEEE, 2012, pp. 1287--1294.

\bibitem{sutton1998reinforcement}
R.~S. Sutton and A.~G. Barto, \emph{Reinforcement learning: An
  introduction}.\hskip 1em plus 0.5em minus 0.4em\relax MIT press Cambridge,
  1998, vol.~1.

\bibitem{Dutreilh11}
X.~Dutreilh, S.~Kirgizov, O.~Melekhova, J.~Malenfant, N.~Rivierre, and
  I.~Truck, ``Using reinforcement learning for autonomic resource allocation in
  clouds: towards a fully automated workflow,'' in \emph{ICAS 2011, The Seventh
  International Conference on Autonomic and Autonomous Systems}, 2011, pp.
  67--74.

\bibitem{Barrett2013}
E.~Barrett, E.~Howley, and J.~Duggan, ``Applying reinforcement learning towards
  automating resource allocation and application scalability in the cloud,''
  \emph{Concurrency and Computation: Practice and Experience}, vol.~25, no.~12,
  pp. 1656--1674, 2013.

\bibitem{Galstyan04}
A.~Galstyan, K.~Czajkowski, and K.~Lerman, ``Resource allocation in the grid
  using reinforcement learning,'' in \emph{Proceedings of the Third
  International Joint Conference on Autonomous Agents and Multiagent
  Systems-Volume 3}.\hskip 1em plus 0.5em minus 0.4em\relax IEEE Computer
  Society, 2004, pp. 1314--1315.

\bibitem{tesauro2007managing}
G.~Tesauro, R.~Das, H.~Chan, J.~Kephart, D.~Levine, F.~Rawson, and C.~Lefurgy,
  ``Managing power consumption and performance of computing systems using
  reinforcement learning,'' in \emph{Advances in Neural Information Processing
  Systems}, 2007, pp. 1497--1504.

\bibitem{lin2016reinforcement}
X.~Lin, Y.~Wang, and M.~Pedram, ``A reinforcement learning-based power
  management framework for green computing data centers,'' in \emph{2016 IEEE
  International Conference on Cloud Engineering (IC2E)}.\hskip 1em plus 0.5em
  minus 0.4em\relax IEEE, 2016, pp. 135--138.

\bibitem{benini2000survey}
L.~Benini, A.~Bogliolo, and G.~De~Micheli, ``A survey of design techniques for
  system-level dynamic power management,'' \emph{IEEE transactions on very
  large scale integration (VLSI) systems}, vol.~8, no.~3, pp. 299--316, 2000.

\bibitem{dhiman2006dynamic}
G.~Dhiman and T.~S. Rosing, ``Dynamic power management using machine
  learning,'' in \emph{Proceedings of the 2006 IEEE/ACM international
  conference on Computer-aided design}.\hskip 1em plus 0.5em minus 0.4em\relax
  ACM, 2006, pp. 747--754.

\bibitem{jung2007dynamic}
H.~Jung and M.~Pedram, ``Dynamic power management under uncertain
  information,'' in \emph{2007 Design, Automation \& Test in Europe Conference
  \& Exhibition}.\hskip 1em plus 0.5em minus 0.4em\relax IEEE, 2007, pp. 1--6.

\bibitem{silver2016mastering}
D.~Silver, A.~Huang, C.~J. Maddison, A.~Guez, L.~Sifre, G.~Van Den~Driessche,
  J.~Schrittwieser, I.~Antonoglou, V.~Panneershelvam, M.~Lanctot \emph{et~al.},
  ``Mastering the game of go with deep neural networks and tree search,''
  \emph{Nature}, vol. 529, no. 7587, pp. 484--489, 2016.

\bibitem{mnih2013playing}
V.~Mnih, K.~Kavukcuoglu, D.~Silver, A.~Graves, I.~Antonoglou, D.~Wierstra, and
  M.~Riedmiller, ``Playing atari with deep reinforcement learning,''
  \emph{arXiv preprint arXiv:1312.5602}, 2013.

\bibitem{bengio2009learning}
Y.~Bengio, ``Learning deep architectures for ai,'' \emph{Foundations and
  trends{\textregistered} in Machine Learning}, vol.~2, no.~1, pp. 1--127,
  2009.

\bibitem{hochreiter1997long}
S.~Hochreiter and J.~Schmidhuber, ``Long short-term memory,'' \emph{Neural
  computation}, vol.~9, no.~8, pp. 1735--1780, 1997.

\bibitem{clusterdata:Reiss2011}
C.~Reiss, J.~Wilkes, and J.~L. Hellerstein, ``{Google} cluster-usage traces:
  format + schema,'' Google Inc., Mountain View, CA, USA, Technical Report,
  Nov. 2011, revised 2012.03.20. Posted at
  url{http://code.google.com/p/googleclusterdata/wiki/TraceVersion2}.

\bibitem{watkins1992q}
C.~J. Watkins and P.~Dayan, ``Q-learning,'' \emph{Machine learning}, vol.~8,
  no. 3-4, pp. 279--292, 1992.

\bibitem{duff1995reinforcement}
S.~J. Duff and O.~Bradtke~Michael, ``Reinforcement learning methods for
  continuous-time markov decision problems,'' \emph{Adv Neural Inf Process
  Syst}, vol.~7, p. 393, 1995.

\bibitem{wang2011deriving}
Y.~Wang, Q.~Xie, A.~Ammari, and M.~Pedram, ``Deriving a near-optimal power
  management policy using model-free reinforcement learning and bayesian
  classification,'' in \emph{Proceedings of the 48th Design Automation
  Conference}.\hskip 1em plus 0.5em minus 0.4em\relax ACM, 2011, pp. 41--46.

\bibitem{sutton1996generalization}
R.~S. Sutton, ``Generalization in reinforcement learning: Successful examples
  using sparse coarse coding,'' \emph{Advances in neural information processing
  systems}, pp. 1038--1044, 1996.

\bibitem{fan2007power}
X.~Fan, W.-D. Weber, and L.~A. Barroso, ``Power provisioning for a
  warehouse-sized computer,'' vol.~35, no.~2, pp. 13--23, 2007.

\bibitem{meisner2009powernap}
D.~Meisner, B.~T. Gold, and T.~F. Wenisch, ``Powernap: eliminating server idle
  power,'' in \emph{ACM Sigplan Notices}, vol.~44, no.~3.\hskip 1em plus 0.5em
  minus 0.4em\relax ACM, 2009, pp. 205--216.

\bibitem{connor1994recurrent}
J.~T. Connor, R.~D. Martin, and L.~E. Atlas, ``Recurrent neural networks and
  robust time series prediction,'' \emph{IEEE transactions on neural networks},
  vol.~5, no.~2, pp. 240--254, 1994.

\bibitem{tan2009adaptive}
Y.~Tan, W.~Liu, and Q.~Qiu, ``Adaptive power management using reinforcement
  learning,'' in \emph{Proceedings of the 2009 International Conference on
  Computer-Aided Design}.\hskip 1em plus 0.5em minus 0.4em\relax ACM, 2009, pp.
  461--467.

\bibitem{mnih2015human}
V.~Mnih, K.~Kavukcuoglu, D.~Silver, A.~A. Rusu, J.~Veness, M.~G. Bellemare,
  A.~Graves, M.~Riedmiller, A.~K. Fidjeland, G.~Ostrovski \emph{et~al.},
  ``Human-level control through deep reinforcement learning,'' \emph{Nature},
  vol. 518, no. 7540, pp. 529--533, 2015.

\bibitem{Rao2009}
J.~Rao, X.~Bu, C.-Z. Xu, L.~Wang, and G.~Yin, ``Vconf: a reinforcement learning
  approach to virtual machines auto-configuration,'' in \emph{Proceedings of
  the 6th international conference on Autonomic computing}.\hskip 1em plus
  0.5em minus 0.4em\relax ACM, 2009, pp. 137--146.

\bibitem{kingma2014adam}
D.~Kingma and J.~Ba, ``Adam: A method for stochastic optimization,''
  \emph{arXiv preprint arXiv:1412.6980}, 2014.

\bibitem{gross2008fundamentals}
D.~Gross, \emph{Fundamentals of queueing theory}.\hskip 1em plus 0.5em minus
  0.4em\relax John Wiley \& Sons, 2008.

\bibitem{lecun1995convolutional}
Y.~LeCun and Y.~Bengio, ``Convolutional networks for images, speech, and time
  series,'' \emph{The handbook of brain theory and neural networks}, vol. 3361,
  no.~10, p. 1995, 1995.

\bibitem{Srivastava:1996:PSS:231913.231922}
\BIBentryALTinterwordspacing
M.~B. Srivastava, A.~P. Chandrakasan, and R.~W. Brodersen, ``Predictive system
  shutdown and other architectural techniques for energy efficient programmable
  computation,'' \emph{IEEE Trans. Very Large Scale Integr. Syst.}, vol.~4,
  no.~1, pp. 42--55, Mar. 1996. [Online]. Available:
  \url{http://dx.doi.org/10.1109/92.486080}
\BIBentrySTDinterwordspacing

\bibitem{643266}
C.-H. Hwang and A.~C.~H. Wu, ``A predictive system shutdown method for energy
  saving of event-driven computation,'' in \emph{1997 Proceedings of IEEE
  International Conference on Computer Aided Design (ICCAD)}, Nov 1997, pp.
  28--32.

\bibitem{kalchbrenner2015grid}
N.~Kalchbrenner, I.~Danihelka, and A.~Graves, ``Grid long short-term memory,''
  \emph{arXiv preprint arXiv:1507.01526}, 2015.

\bibitem{DBLP:journals/corr/KingmaB14}
\BIBentryALTinterwordspacing
D.~P. Kingma and J.~Ba, ``Adam: {A} method for stochastic optimization,''
  \emph{CoRR}, vol. abs/1412.6980, 2014. [Online]. Available:
  \url{http://arxiv.org/abs/1412.6980}
\BIBentrySTDinterwordspacing

\bibitem{931003}
T.~Simunic, L.~Benini, P.~Glynn, and G.~D. Micheli, ``Event-driven power
  management,'' \emph{IEEE Transactions on Computer-Aided Design of Integrated
  Circuits and Systems}, vol.~20, no.~7, pp. 840--857, Jul 2001.

\bibitem{qiu2007stochastic}
Q.~Qiu, Y.~Tan, and Q.~Wu, ``Stochastic modeling and optimization for robust
  power management in a partially observable system,'' in \emph{Proceedings of
  the conference on Design, automation and test in Europe}.\hskip 1em plus
  0.5em minus 0.4em\relax EDA Consortium, 2007, pp. 779--784.

\bibitem{little2008little}
J.~D. Little and S.~C. Graves, ``Little's law,'' in \emph{Building
  intuition}.\hskip 1em plus 0.5em minus 0.4em\relax Springer, 2008, pp.
  81--100.

\bibitem{wang2015dueling}
Z.~Wang, N.~de~Freitas, and M.~Lanctot, ``Dueling network architectures for
  deep reinforcement learning,'' \emph{arXiv preprint arXiv:1511.06581}, 2015.

\end{thebibliography}

\end{document}